\documentclass[sigconf, screen, nonacm]{acmart}

\setcopyright{none}
\usepackage{siunitx}

\usepackage{braket}
\providecommand{\vk}[1]{}
\providecommand{\da}[1]{}

\usepackage{blindtext}
\usepackage[utf8]{inputenc}
\usepackage{epsfig,xspace,url,xcolor}
\usepackage{amsmath}
\usepackage{mathtools}
\usepackage{textcomp}
\usepackage{amsthm}
\usepackage{tikz}
\usepackage{pgfplots}
\usepackage{placeins}
\pgfplotsset{compat=1.18}
\usetikzlibrary{patterns}
\usepgfplotslibrary{groupplots}
\providecommand{\QuantumPlotCompat}{1.18}
\pgfplotsset{compat=\QuantumPlotCompat}
\definecolor{evalfixedclos}{HTML}{8C8C8C}
\definecolor{evalSwitchQNet}{HTML}{E69F00}
\definecolor{evalred}{HTML}{D55E00}
\definecolor{evalredQGP}{HTML}{0072B2}

\usetikzlibrary{patterns}
\usepgfplotslibrary{groupplots}
\definecolor{qvcolor0}{HTML}{8C8C8C}
\definecolor{qvcolor1}{HTML}{CC79A7}
\definecolor{qvcolor2}{HTML}{D55E00}
\definecolor{qvcolor3}{HTML}{0072B2}
\definecolor{qvcolor4}{HTML}{E69F00}
\definecolor{qvcolor5}{HTML}{009E73}

\usetikzlibrary{positioning,calc}
\usepackage{caption}
\usepackage{graphicx}
\usepackage[inline]{enumitem}
\usepackage{tikz}
\usetikzlibrary{arrows.meta, positioning, shapes.geometric, shadows, patterns}
\usepackage{thmtools}
\usepackage{thm-restate}
\usepackage{scalerel}
\usepackage[ruled, lined, linesnumbered, commentsnumbered, noend]{algorithm2e}
\usepackage[noend]{algpseudocode}
\usepackage{graphicx}
\usepackage{blindtext}
\usepackage{url}
\usepackage{subcaption}
\usepackage{mathtools}
\usepackage{xcolor}
\usepackage{color, colortbl}
\usepackage{tablefootnote}
\usepackage[subtle]{savetrees}
\usepackage{graphbox}
\usepackage{url}

\usepackage[normalem]{ulem}
\usepackage{wrapfig}
\usepackage{wrapstuff}
\usepackage[framemethod=TikZ]{mdframed}

\usepackage{enumitem}

\colorlet{rowHighlight}{green!10}
\definecolor{darkgreen}{rgb}{0.01, 0.75, 0.24}
\definecolor{darkcyan}{rgb}{0.0, 0.55, 0.55}
\definecolor{mycolor}{rgb}{0.01, 0.75, 0.24}
\definecolor{takeawaycolor}{rgb}{0.75, 0, 0}
\definecolor{mygray}{gray}{0.9}
\definecolor{myred}{rgb}{0.75, 0, 0}

\definecolor{alg1base}{rgb}{0.957, 1, 0.961}
\colorlet{alg1}{alg1base!90!white}
\definecolor{alg2base}{rgb}{0.957, 0.988, 1}
\colorlet{alg2}{alg2base!90!white}
\definecolor{alg3base}{rgb}{0.824, 0.914, 1}
\colorlet{alg3}{alg3base!90!white}

\mathchardef\hyphen="2D

\newtheoremstyle{sltheorem}
{}                %
{}                %
{\slshape}        %
{}                %
{\bfseries}       %
{.}               %
{ }               %
{}                %
\theoremstyle{sltheorem}

\newmdenv[
  backgroundcolor=gray!20,
  linecolor=gray,
  linewidth=1pt,
  roundcorner=5pt,
  skipabove=6pt,
  skipbelow=6pt,
  innertopmargin=1pt,
  innerbottommargin=4pt,
  innerleftmargin=6pt,   %
  innerrightmargin=6pt   %
]{graybox}

\newcommand{\myitem}[1]{\vspace{0.6mm}\noindent\textbf{#1}}

\newcommand{\remove}[1]{}

\newcommand{\first}{\emph{(i)}\xspace}
\newcommand{\second}{\emph{(ii)}\xspace}
\newcommand{\third}{\emph{(iii)}\xspace}

\usepackage{xcolor}
\definecolor{light-gray}{gray}{0.9}
\usepackage{tcolorbox}

\newcommand{\cref}[1]{Chapter~\ref{#1}}

\DeclareSymbolFont{extraup}{U}{zavm}{m}{n}
\DeclareMathSymbol{\varheart}{\mathalpha}{extraup}{86}
\DeclareMathSymbol{\vardiamond}{\mathalpha}{extraup}{87}

\newcommand{\takeaway}[1]{{\textcolor{black}{$\blacksquare$ \textbf{\textit{Takeaway:}}} \textit{#1}}}

\newcommand{\name}{\textsc{Schism}\xspace}

\begin{document}

\sloppy

\title{Schism: Reconfigurable Scale-Up Domains\\for Distributed Quantum Computing}

\author{Daniel Alexander}
\affiliation{
  \institution{Purdue University}
}
\author{Deeksha P Rao}
\affiliation{
  \institution{Purdue University}
}
\author{Nilesh Vyas}
\affiliation{
  \institution{Airbus Central R\&T}
}
\author{Vamsi Addanki}
\affiliation{
  \institution{Purdue University}
}

\renewcommand{\shortauthors}{Daniel Alexander, Deeksha P Rao, Nilesh Vyas, Vamsi Addanki}

\begin{abstract}
Switch-based quantum datacenters connect quantum processing units (QPUs) through photonic switches and shared Bell-state measurement (BSM) devices. Sharing BSMs saves hardware, but fixed connectivity constrains access to these resources. Local communication can overload one rack's BSMs while others remain idle; reaching those idle resources incurs additional optical loss and retries through switches and quantum frequency converters (QFCs). Programs remain locked into this locality-contention tradeoff as their communication patterns change.

We propose \name, an architecture that lets programs dynamically change which QPUs share each BSM pool. \name places BSM pools at aggregation switches and QFCs on their core-facing uplinks. Top-of-rack switches establish reconfigurable native-wavelength circuits from QPUs to these pools, forming scale-up domains bounded by the optical loss budget. Reconfiguration reshapes resource sharing across racks without moving QPUs or migrating quantum states. Our key insight is that QPU grouping jointly determines photon-path loss and contention for BSM resources: locality alone cannot identify the best grouping, and a grouping pays off only if its savings outweigh the delay of reconfiguring to it. We develop an optimization framework that combines graph partitioning with resource-constrained completion-time estimates and dynamic programming to select \emph{how} QPUs share BSM pools and \emph{when} those groupings reconfigure, between or within circuit phases.
Our evaluation using SeQUeNCe with trapped-ion parameters shows that \name achieves communication speedups of $4.9$--$7.3\times$ across 11 workloads on 32 QPUs compared to Fixed Clos connectivity. For 192-qubit QFT, QAOA, and Grover, the speedups are $4.87\times$, $6.66\times$, and $6.61\times$, respectively, and $6.73\times$, $8.26\times$, and $6.22\times$ compared to SwitchQNet. \name also outperforms a hypothetical network-wide switch by $1.35$--$1.38\times$ on the eight workloads whose phases fit in a pool, and trails it by $1.1$--$1.75\times$ on the other three. The benefits grow with conversion loss: increasing loss per QFC from 0.5 to 5~dB increases the mean speedup compared to Fixed Clos from $2.8\times$ to $20\times$.

\end{abstract}

\maketitle

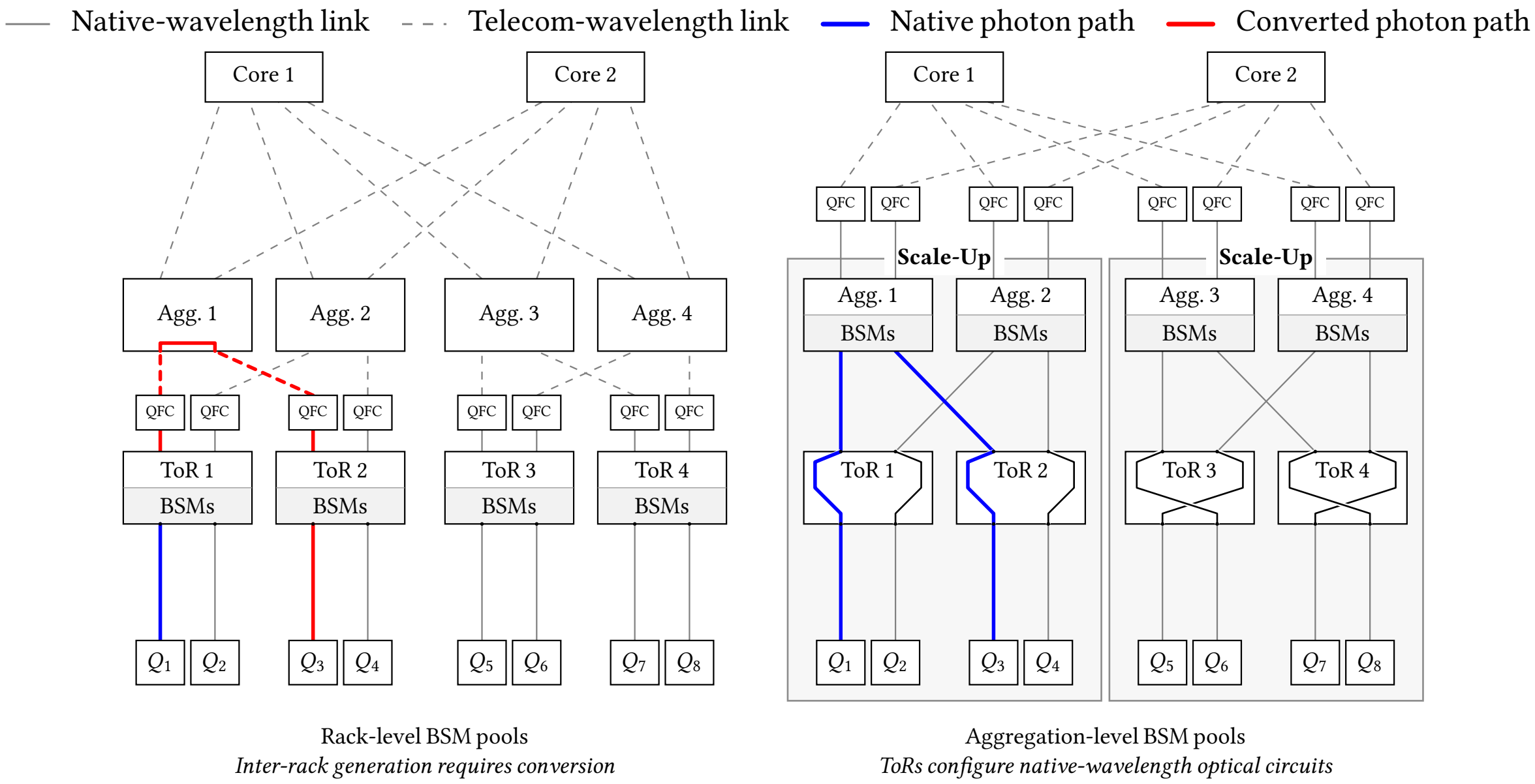
\begin{figure*}[t]
\centering
\begingroup
\captionsetup[subfigure]{font=normalsize,position=bottom,skip=5pt}

\tikzset{
  reddevice/.style={
    draw=black, fill=white, line width=0.55pt,
    inner sep=2pt, align=center
  },
  redpod/.style={
    draw=black!45, fill=black!3,
    line width=0.65pt
  },
  redfiber/.style={draw=gray, line width=0.55pt},
  redcircuit/.style={draw=black, line width=0.65pt, line join=round},
  redtelecom/.style={redfiber, dashed},
  redlocal/.style={
    draw=blue, line width=1.35pt,
    line cap=round, line join=round
  },
  redremote/.style={
    draw=red, line width=1.35pt,
    line cap=round, line join=round
  }
}

\newcommand{\redpodpanel}[1]{%
\begin{tikzpicture}[
  x=0.123\linewidth,
  y=0.123\linewidth,
  font=\normalsize
]
  \path[use as bounding box] (0,-0.45) rectangle (8,8.90);

  \ifnum#1=1
    \foreach \p/\left/\right/\middle in {
      1/0.05/3.95/2,
      2/4.05/7.95/6
    }{
      \draw[redpod] (\left,0.75) rectangle (\right,6.25);
      \node[fill=white,inner xsep=5pt,inner ysep=2pt,
            font=\normalsize\bfseries]
        at (\middle,6.25) {Scale-Up};
    }
  \fi

  \foreach \c/\x in {1/2,2/6}{
    \foreach \p/\a/\dx in {1/1/-0.54,1/2/-0.18,2/1/0.18,2/2/0.54}{
      \coordinate (core-\c-\p-\a) at (\x+\dx,8.20);
    }
  }

  \foreach \p/\shift in {1/0,2/4}{
    \foreach \a/\localx in {1/1.05,2/2.95}{
      \pgfmathsetmacro{\x}{\shift+\localx}

      \foreach \t/\dx in {1/-0.34,2/0.34}{
        \coordinate (agg-down-\p-\a-\t) at (\x+\dx,5.10);
        \coordinate (agg-up-\p-\a-\t) at (\x+\dx,6.00);
        \coordinate (agg-qfc-bottom-\p-\a-\t) at (\x+\dx,6.72);
        \coordinate (agg-qfc-top-\p-\a-\t) at (\x+\dx,7.14);

        \ifnum#1=0
          \draw[redtelecom]
            (agg-up-\p-\a-\t) -- (core-\t-\p-\a);
        \else
          \draw[redfiber]
            (agg-up-\p-\a-\t) -- (agg-qfc-bottom-\p-\a-\t);
          \draw[redtelecom]
            (agg-qfc-top-\p-\a-\t) -- (core-\t-\p-\a);
        \fi
      }
    }

    \foreach \t/\localx in {1/1.05,2/2.95}{
      \pgfmathsetmacro{\x}{\shift+\localx}
      \foreach \a/\dx in {1/-0.34,2/0.34}{
        \coordinate (tor-up-\p-\t-\a) at (\x+\dx,3.85);
        \coordinate (tor-qfc-bottom-\p-\t-\a) at (\x+\dx,4.12);
        \coordinate (tor-qfc-top-\p-\t-\a) at (\x+\dx,4.55);
        \ifnum#1=0
          \draw[redfiber]
            (tor-up-\p-\t-\a) -- (tor-qfc-bottom-\p-\t-\a);
          \draw[redtelecom]
            (tor-qfc-top-\p-\t-\a) -- (agg-down-\p-\a-\t);
        \else
          \draw[redfiber]
            (tor-up-\p-\t-\a) -- (agg-down-\p-\a-\t);
        \fi
      }
    }
  }

  \ifnum#1=0
    \foreach \t in {1,2}{
      \draw[redremote]
        (tor-up-1-\t-1) -- (tor-qfc-bottom-1-\t-1);
      \draw[redremote,dashed]
        (tor-qfc-top-1-\t-1) -- (agg-down-1-1-\t);
    }
  \else
    \draw[redlocal]
      (tor-up-1-1-1) -- (agg-down-1-1-1);
    \draw[redlocal]
      (tor-up-1-2-1) -- (agg-down-1-1-2);
  \fi

  \foreach \c/\x in {1/2,2/6}{
    \draw[reddevice] (\x-0.73,8.20) rectangle (\x+0.73,8.82);
    \node at (\x,8.55) {Core \c};
  }

  \foreach \p/\shift in {1/0,2/4}{
    \foreach \a/\localx in {1/1.05,2/2.95}{
      \pgfmathsetmacro{\x}{\shift+\localx}
      \pgfmathtruncatemacro{\n}{2*(\p-1)+\a}
      \draw[reddevice] (\x-0.80,5.10) rectangle (\x+0.80,6.00);
      \ifnum#1=0
        \node at (\x,5.55) {Agg.\ \n};
      \else
        \fill[gray!10] (\x-0.79,5.11) rectangle (\x+0.79,5.54);
        \draw[black!40,line width=0.35pt]
          (\x-0.80,5.55) -- (\x+0.80,5.55);
        \node at (\x,5.775) {Agg.\ \n};
        \node at (\x,5.325) {BSMs};
        \foreach \dx in {-0.34,0.34}{
          \draw[reddevice] (\x+\dx-0.30,6.72) rectangle (\x+\dx+0.30,7.14);
          \node[inner sep=1pt,font=\scriptsize] at (\x+\dx,6.93) {QFC};
        }
      \fi
    }

    \foreach \t/\localx in {1/1.05,2/2.95}{
      \pgfmathsetmacro{\x}{\shift+\localx}
      \pgfmathtruncatemacro{\n}{2*(\p-1)+\t}
      \draw[reddevice] (\x-0.80,2.95) rectangle (\x+0.80,3.85);
      \node at (\x,3.625) {ToR \n};
      \ifnum#1=0
        \fill[gray!10] (\x-0.79,2.96) rectangle (\x+0.79,3.39);
        \draw[black!40,line width=0.35pt]
          (\x-0.80,3.40) -- (\x+0.80,3.40);
        \node at (\x,3.175) {BSMs};
        \foreach \dx in {-0.34,0.34}{
          \draw[reddevice] (\x+\dx-0.30,4.12) rectangle (\x+\dx+0.30,4.55);
          \node[inner sep=1pt,font=\scriptsize] at (\x+\dx,4.335) {QFC};
        }
      \else
        \ifnum\p=1
          \def\redportmap{1/1,2/2}
        \else
          \def\redportmap{1/2,2/1}
        \fi
        \foreach \inputport/\outputport in \redportmap{
          \pgfmathsetmacro{\inputx}{\x+0.34*(2*\inputport-3)}
          \pgfmathsetmacro{\outputx}{\x+0.34*(2*\outputport-3)}
          \pgfmathsetmacro{\sidex}{\x+0.66*(2*\outputport-3)}
          \draw[redcircuit,preaction={draw=white,line width=1.8pt}]
            (\inputx,2.95) -- (\inputx,3.08) -- (\sidex,3.40)
            -- (\sidex,3.72) -- (\outputx,3.85);
        }
        \ifnum\p=1
          \draw[redlocal]
            (\x-0.34,2.95) -- (\x-0.34,3.08) -- (\x-0.66,3.40)
            -- (\x-0.66,3.72) -- (tor-up-\p-\t-1);
        \fi
        \foreach \a in {1,2}{
          \fill[black] (tor-up-\p-\t-\a) circle (0.022);
        }
      \fi
    }
  }
  \ifnum#1=0
    \draw[redremote]
      (agg-down-1-1-1) -- ++(0,0.10) -| (agg-down-1-1-2);
  \fi

  \foreach \i/\x in {
    1/0.71,2/1.39,3/2.61,4/3.29,
    5/4.71,6/5.39,7/6.61,8/7.29
  }{
    \coordinate (tor-port-\i) at (\x,2.95);
    \coordinate (qpu-port-\i) at (\x,1.50);
    \draw[redfiber] (qpu-port-\i) -- (tor-port-\i);
  }
  \draw[redlocal] (qpu-port-1) -- (tor-port-1);
  \ifnum#1=0
    \draw[redremote] (qpu-port-3) -- (tor-port-3);
  \else
    \draw[redlocal] (qpu-port-3) -- (tor-port-3);
  \fi

  \foreach \i/\x in {
    1/0.71,2/1.39,3/2.61,4/3.29,
    5/4.71,6/5.39,7/6.61,8/7.29
  }{
    \draw[reddevice] (\x-0.30,0.95) rectangle (\x+0.30,1.50);
    \node at (\x,1.225) {$Q_{\i}$};
    \fill[black] (tor-port-\i) circle (0.022);
  }

  \ifnum#1=0
    \node[text=black,align=center] at (4,0.10)
      {Rack-level BSM pools\\\textit{Inter-rack generation requires conversion}};
  \else
    \node[text=black,align=center] at (4,0.10)
      {Aggregation-level BSM pools\\\textit{ToRs configure native-wavelength optical circuits}};
  \fi
\end{tikzpicture}%
}

\setlength{\tabcolsep}{3pt}
\begin{tabular}{@{}ll@{\quad}ll@{\quad}ll@{\quad}ll@{}}
  \tikz[baseline=-0.5ex]\draw[redfiber] (0,0) -- (0.45,0);
  & Native-wavelength link
  & \tikz[baseline=-0.5ex]\draw[redtelecom] (0,0) -- (0.45,0);
  & Telecom-wavelength link
  & \tikz[baseline=-0.5ex]\draw[redlocal] (0,0) -- (0.45,0);
  & Native photon path
  & \tikz[baseline=-0.5ex]\draw[redremote] (0,0) -- (0.45,0);
  & Converted photon path
\end{tabular}

\resizebox{0.76\textwidth}{!}{%
\begin{minipage}{\textwidth}
\centering
\begin{subfigure}[t]{0.49\textwidth}
  \centering
  \redpodpanel{0}
  \vspace{-3mm}
  \caption{Prior work (e.g., SwitchQNet): Fixed QPU-BSM sharing}
  \label{fig:architecture-fixed}
\end{subfigure}
\hfill
\begin{subfigure}[t]{0.49\textwidth}
  \centering
  \redpodpanel{1}
  \vspace{-3mm}
  \caption{\name: Reconfigurable QPU-BSM sharing}
  \label{fig:architecture-red}
\end{subfigure}
\end{minipage}%
}

\endgroup
\vspace{-2mm}
\caption{\name enables programmable QPU-BSM sharing through optical circuits at the ToRs. \textit{(a)} Prior designs (e.g., SwitchQNet) place BSMs and QFCs at the ToRs. To generate a Bell pair between $Q_1$ and $Q_3$, the photon from $Q_3$ traverses two QFCs and an aggregation switch before reaching ToR~1's BSM. \textit{(b)} \name places BSM pools at aggregation switches and QFCs on their core-facing uplinks. The ToRs connect both QPUs to the same BSM pool through native-wavelength circuits, avoiding conversion. Reconfiguring the circuits changes which QPUs share each pool within a scale-up domain, enabling reconfigurable BSM sharing without moving QPUs or stored quantum states. 
The optical loss budget determines the scale-up extent: low-loss ToR switches enable BSM sharing across multiple racks, potentially at the cost of higher reconfiguration delay, while higher loss can restrict it to a single rack. Exploiting this reach requires choosing \emph{when} and \emph{how} to reconfigure circuits, balancing BSM and uplink contention against optical loss and reconfiguration delay.}
\label{fig:architecture}
\vspace{-3mm}
\end{figure*}

\vspace{-2pt}
\section{Introduction}
\label{sec:introduction}

Quantum computing offers new approaches to factoring large integers~\cite{doi:10.1137/S0097539795293172, Gidney2021howtofactorbit, 365700}, simulating quantum systems~\cite{Daley2022,c7bc8b77-97a9-3229-bba6-29cbf867d990,RevModPhys.86.153, Montanaro2016}, and combinatorial optimization~\cite{farhi2014qaoa, Abbas2024}. Scaling these capabilities beyond a single quantum processing unit (QPU) motivates quantum datacenters, where interconnected processors execute quantum programs collectively \cite{Main2025, 10521772,cuomo2020towards, CALEFFI2024110672, 10.1145/3674151}. Operations spanning QPUs consume shared Bell pairs, making entanglement generation essential to distributed execution. Yet establishing these pairs is costly: optical loss and probabilistic generation require repeated attempts, while contention for shared resources delays communication. Adding processors expands computational capacity, but realizing that capacity requires an architecture that enables efficient entanglement generation between them~\cite{Cacciapuoti_2026, BIRLICA2026108597, PRXQuantum, monroe2014large}.

Quantum datacenter architectures range from direct QPU-to-QPU interconnects to switch-based systems, such as SwitchQNet~\cite{zhang2025switchqnet}, inspired by classical leaf-spine and Clos 
topologies~\cite{shapourian2025quantum, 10.1145/1402958.1402967}. In switch-based systems, photonic switches route photons from QPU interfaces to shared Bell-state measurement (BSM) devices~\cite{shapourian2025quantum, tekleab2026distributedswitchingprotocolquantum, Koyama_2024, Mazza_2025, 9796814}.
A successful measurement can establish a Bell pair between qubits at the source QPUs. 

Switch-based architectures~\cite{earl2022architecturefirstgenerationcommercialquantum, blanco2025physicallogicalgraphstatebasedconnectivity} offer two main benefits. First, they allow different QPU pairs to share BSM pools, reducing hardware duplication and allowing provisioning to reflect concurrent demand. Second, quantum frequency converters (QFCs) extend communication beyond the practical reach of native-wavelength links by translating photons to telecom wavelengths for lower-loss fiber transmission. Placing QFCs on switch uplinks provides this reach without requiring a dedicated converter at every QPU~\cite{zhang2025switchqnet}.

These benefits raise two architectural questions: which QPUs should share each BSM pool, and where should wavelength conversion occur? In designs with BSMs and QFCs at top-of-rack (ToR) switches, physical connectivity fixes which QPUs have direct, native-wavelength access to each pool. Reaching another ToR's pool requires additional switches and conversion. In Figure~\ref{fig:architecture-fixed}, the photon from $Q_3$ traverses two QFCs and an aggregation switch to reach ToR~1's BSM and generate a Bell pair with $Q_1$.

Grouping frequently communicating QPUs around one BSM pool appears to offer a straightforward way to avoid these lossy paths~\cite{9334411, 10628181}. However, these QPUs also compete for the same limited resources. When several pairs request entanglement simultaneously, only a subset can proceed; the others must wait for a BSM to become available. Generation retries prolong resource occupancy, increasing this waiting time. The time spent waiting for BSM resources can outweigh the benefit of shorter photon paths.

Distributing requests across BSM pools can relieve contention by allowing more entanglement-generation attempts to proceed concurrently~\cite{pouryousef2026benchmarking}. Reaching those pools, however, requires additional optical paths through switches and potentially wavelength converters. The added loss increases generation retries, while contention for shared uplinks can further delay communication. Access to more BSMs offers greater parallelism, but introduces additional optical loss and consumes shared network resources.

The preferred grouping depends on the program and can change as its circuit progresses. One phase may concentrate requests among a few QPUs, while another spreads communication across many pairs. Qubit placement can reduce communication, and scheduling can coordinate resource use, but neither changes the network's physical QPU-to-BSM connectivity. A fixed architecture locks programs into a rigid tradeoff: BSM capacity may be available, yet costly for the QPUs that need it to reach.

We propose \name, Reconfigurable Scale-Up Domains, an architecture that enables programmable sharing of BSM resources. Our design builds on two observations. First, a rack boundary is not a wavelength-conversion boundary. Native-wavelength paths can span multiple racks within the available optical loss budget, yet QFCs on every ToR uplink (e.g., SwitchQNet) force inter-ToR communication through conversion. Second, QPUs need not share BSM resources according to their physical rack placement. Optical circuit switching can determine which QPUs access each pool without moving QPUs or migrating quantum states.

As Figure~\ref{fig:architecture-red} shows, \name uses ToRs as native-wavelength optical circuit switches and places BSM pools at aggregation switches. Each ToR establishes circuits that assign QPUs to selected BSM pools. QPUs assigned to the same pool share local entanglement-generation resources and can establish Bell pairs without frequency conversion. A scale-up domain comprises racks     connected to a common set of aggregation switches that host BSM resources: its ToRs can change which QPUs share each BSM pool. The scale-up extent is bounded by the native-wavelength optical loss budget, requiring low insertion loss at the ToR switches. QPUs assigned to different pools may still communicate through QFC-equipped uplinks and the core, even within the same~scale-up domain. 

The central decision becomes \emph{how} QPUs should share BSM pools and \emph{when} that sharing should change. Our key insight is that grouping jointly determines photon paths and contention for resources, while its benefit depends on how long the program can exploit it. Optimizing each phase independently can sacrifice these benefits to frequent reconfiguration. We instead optimize the configuration sequence across circuit phases; with millisecond switches, the compiler may also regroup within a phase to keep a cross-pool pair in one pool. We use graph partitioning on QPU interaction graphs weighted by Bell-pair requests to identify communicating groups, then estimate their completion times under optical loss and BSM and link contention. Dynamic programming selects a sequence that minimizes estimated total communication time, including reconfiguration delay. This framework retains a configuration when switching offers insufficient benefit and changes it when the program's communication savings justify the transition.

We evaluate \name using SeQUeNCe with trapped-ion parameters~\cite{stephenson2020high} on MQT Bench workloads~\cite{quetschlich2023mqtbench} and networks with 16 to 64 QPUs. Across all 11 workloads on 32 QPUs, \name achieves communication speedups of $4.9$--$7.3\times$ (mean $6.4\times$) compared to Fixed Clos connectivity and $3.6$--$8.3\times$ compared to SwitchQNet~\cite{zhang2025switchqnet}. \name eliminates all 27,705 and 28,916 Bell-pair generations that traverse QFCs and the core under Fixed Clos for 192-qubit QFT and Grover, respectively. Regrouping accounts for up to about 2.5\% of total communication time. These benefits extend to larger systems: for 384-qubit QAOA on 64 QPUs, \name achieves speedups of $7.78\times$ compared to Fixed Clos connectivity and $10.63\times$ compared to SwitchQNet.

In summary, our key contributions are as follows:
\begin{itemize}[label=\small{\textcolor{black}{$\blacksquare$}}, leftmargin=*]
    \item We identify the tradeoff between communication locality and BSM contention in fixed quantum datacenter architectures, showing why minimizing photon-path loss alone is insufficient to minimize communication time.

    \item We propose \name, an architecture that combines aggregation-level BSM pools with native-wavelength circuit switching at the ToRs to make QPU-to-BSM sharing programmable across racks without migrating quantum states.

    \item We develop a program-aware optimization framework that combines graph partitioning, resource-constrained completion-time estimation, and dynamic programming to jointly select QPU groupings and their configuration sequence across circuit phases.

    \item We evaluate \name using SeQUeNCe against Fixed Clos connectivity and SwitchQNet, quantifying its benefits and the tradeoffs among path loss, resource contention, and reconfiguration delay.
\end{itemize}

\section{Motivation}
\label{sec:motivation}

We first describe the architecture and assumptions (\S\ref{sec:architecture-assumptions}), then show the benefits and limitations of scheduling under fixed connectivity (\S\ref{sec:rigid-tradeoff}). We next discuss how reconfigurable BSM sharing can reduce contention and photon-path loss, and when reconfiguration is worthwhile (\S\ref{sec:reconfigurable-sharing}). Finally, we discuss our design goals and the challenges ahead (\S\ref{sec:design-goals}).

\subsection{Architecture \& Assumptions}
\label{sec:architecture-assumptions}

We consider a datacenter with $n$ QPUs connected by a switch-based photonic interconnect following a Clos topology, widely used in classical datacenters and recently adopted in quantum datacenter designs~\cite{shapourian2025quantum, 10.1145/1402958.1402967}. As Figure~\ref{fig:architecture-fixed} shows, each rack contains $n_u$ QPUs connected to a top-of-rack (ToR) switch, whose uplinks connect to aggregation switches. Core switches interconnect the aggregation switches to provide paths throughout the datacenter.

\myitem{Photonic interconnect:} Photonic circuit switches establish optical paths by directing photons from input ports to configured output ports. Switches may also incorporate Bell-state measurement (BSM) devices and quantum frequency converters (QFCs)~\cite{promponas2026ratefidelitytradeoffsallphotonicmemoryequipped, vanleent2022entangling}. BSMs provide shared measurement resources for entanglement generation: each serves one photon pair at a time, while a pool of BSMs allows multiple pairs to be processed concurrently. QFCs extend communication reach by translating photons between the QPUs' native wavelengths (e.g., near-infrared) and telecom wavelengths (e.g., O-band or C-band) for lower-loss transmission over longer distances~\cite{Rowland_2025}. Where these devices are placed determines which QPUs share each BSM pool and which paths require conversion.

\myitem{Entanglement generation:} We consider two-photon heralded generation, in which each participating QPU prepares entanglement between a stationary communication qubit and an emitted photon. The network routes both photons to a common BSM, where an accepted measurement outcome heralds entanglement between the communication qubits at the source QPUs. The photons are measured, while the resulting Bell pair remains at the QPUs for use by a remote operation. A failed attempt requires the communication qubits to be reset before generation is retried. Successful pairs remain in memory until needed, but storage capacity and decoherence limit how many can be prepared in advance and how long they remain usable~\cite{piparo2026adaptiveresourcememorycontrol, elsayed2026balancingquantummemoriesasymmetric}.

\myitem{Optical loss and generation time:} Each photon accumulates loss from fibers, switches, and any wavelength converters along its path to the BSM, reducing the probability that it arrives for measurement. For photon $i$, the total path loss in dB is
\begin{equation}
L_i =
\sum_{e \in \mathcal{F}_i} \alpha_e d_e
+ \sum_{s \in \mathcal{S}_i} \ell_s
+ \sum_{c \in \mathcal{C}_i} \ell_c,
\label{eq:path_loss}
\end{equation}
where $\mathcal{F}_i$, $\mathcal{S}_i$, and $\mathcal{C}_i$ denote the fibers, switches, and converters along its path. Here, $\alpha_e$ is the fiber attenuation at the transmitted wavelength, $d_e$ is the fiber length, and $\ell_s$ and $\ell_c$ are switch insertion loss and conversion loss, giving a path transmission probability of $10^{-L_i/10}$.

Generation succeeds only if both photons arrive and are detected. Assuming independent emission, transmission, and detection, the per-attempt success probability is
\begin{equation}
p_{\mathrm{gen}}
=
p_{\mathrm{BSM}}
\prod_{i=1}^{2}
\left(
\eta_{\mathrm{emit},i}
\eta_{\mathrm{det},i}
10^{-L_i/10}
\right),
\label{eq:p_gen}
\end{equation}
where $\eta_{\mathrm{emit},i}$ includes emission and collection efficiency, $\eta_{\mathrm{det},i}$ is detection efficiency, and $p_{\mathrm{BSM}}$ is the BSM acceptance probability conditional on both photons being detected. We use measured values for a trapped-ion node (\S\ref{sec:eval-method}); for a given implementation, additional loss on either photon's path reduces the probability of establishing a Bell pair.

For independent attempts with constant success probability, the expected number of attempts is $1/p_{\mathrm{gen}}$. If each attempt takes $\tau_{\mathrm{attempt}}$, including preparation, transmission, heralding, and reset as applicable, the expected generation time is $E[T_{\mathrm{gen}}] = \tau_{\mathrm{attempt}} / p_{\mathrm{gen}}$.  Because the path setup time $t_{setup}$ is incurred only once per routing assignment and the optical paths are reused across all subsequent retries, the total expected service time is:
\begin{equation}
E[T_{\mathrm{service}}] = E[T_{\mathrm{gen}}] + t_{setup} = \frac{\tau_{\mathrm{attempt}}}{p_{\mathrm{gen}}} + t_{setup}.
\end{equation}
Once established, these paths keep the BSM and optical links reserved until generation succeeds, potentially delaying other requests without incurring repeated setup penalties. Pairs generated earlier may wait in memory for other pairs or operations to complete, losing fidelity through decoherence during this wait. Delayed generation can affect both program progress and the fidelity of pairs when they are used.

\medskip
\takeaway{Remote-entanglement completion time depends on both photon-path loss, which increases generation retries, and contention for the BSMs needed to establish each Bell pair.}

\subsection{Limitations of Scheduling}
\label{sec:rigid-tradeoff}

\begin{figure*}[t]
  \centering
  \begin{subfigure}[t]{0.15\textwidth}
    \centering
    \begin{minipage}[c][64mm][c]{\linewidth}
      \centering
      \begingroup
\usetikzlibrary{decorations.pathreplacing}
\newcommand{\redcircuitexample}[1]{%
  \begin{tikzpicture}[x=0.9mm,y=1.5mm,font=\normalsize]
    \path[use as bounding box] (-13,-25) rectangle (15,2);
    \foreach \i in {1,...,8}{
      \ifnum#1=0
        \ifnum\i<8
          \pgfmathsetmacro{\wy}{-3*(\i-1)}
          \pgfmathtruncatemacro{\qid}{\i}
        \else
          \def\wy{-23}
          \def\qid{17}
        \fi
      \else
        \ifnum\i<5
          \pgfmathsetmacro{\wy}{-3*(\i-1)}
          \pgfmathtruncatemacro{\qid}{\i}
        \else
          \pgfmathsetmacro{\wy}{-13-3*(\i-5)}
          \pgfmathtruncatemacro{\qid}{\i+12}
        \fi
      \fi
      \draw[black!55,line width=0.4pt] (0,\wy) -- (15,\wy);
      \node[anchor=east,inner sep=1pt] at (-1,\wy) {$Q_{\qid}$};
    }
    \ifnum#1=0
      \foreach \i in {1,2,3}{
        \pgfmathsetmacro{\cy}{-6*(\i-1)}
        \pgfmathsetmacro{\ty}{\cy-3}
        \draw[line width=0.55pt] (7.5,\cy) -- (7.5,\ty-1);
        \fill (7.5,\cy) circle (0.5mm);
        \draw[line width=0.55pt] (7.5,\ty) circle (0.8mm);
      }
      \draw[line width=0.55pt] (7.5,-18) -- (7.5,-24);
      \fill (7.5,-18) circle (0.5mm);
      \draw[line width=0.55pt] (7.5,-23) circle (0.8mm);
    \else
      \foreach \i in {1,...,4}{
        \pgfmathsetmacro{\gx}{2+3.5*(\i-1)}
        \pgfmathsetmacro{\cy}{-3*(\i-1)}
        \pgfmathsetmacro{\ty}{-13-3*(\i-1)}
        \draw[line width=0.55pt] (\gx,\cy) -- (\gx,\ty-1);
        \fill (\gx,\cy) circle (0.5mm);
        \draw[line width=0.55pt] (\gx,\ty) circle (0.8mm);
      }
    \fi
    \ifnum#1=0
      \draw[decorate,decoration={brace,amplitude=3pt},line width=0.45pt]
        (-7.5,-19.5) -- (-7.5,1.5);
      \node[rotate=90,inner sep=0pt] at (-12,-9) {ToR A};
      \draw[decorate,decoration={brace,amplitude=3pt},line width=0.45pt]
        (-7.5,-24.5) -- (-7.5,-21.5);
      \node[rotate=90,inner sep=0pt] at (-12,-23) {ToR B};
    \else
      \draw[decorate,decoration={brace,amplitude=3pt},line width=0.45pt]
        (-7.5,-10) -- (-7.5,1.5);
      \node[rotate=90,inner sep=0pt] at (-12,-4.25) {ToR A};
      \draw[decorate,decoration={brace,amplitude=3pt},line width=0.45pt]
        (-7.5,-23.5) -- (-7.5,-12);
      \node[rotate=90,inner sep=0pt] at (-12,-17.75) {ToR B};
    \fi
  \end{tikzpicture}%
}
\pgfplotsset{redtradeoff/.style={
  width=\linewidth,height=48mm,
  axis lines=left,xmin=2,xmax=8.3,ymin=0,ymax=1500,
  xtick={2,4,6,8},ytick={0,500,1000,1500},
  scaled y ticks=false,
  ylabel={Bell-pair completion time (ms)},
  tick label style={font=\normalsize},label style={font=\normalsize},
  legend style={draw=none,fill=none,font=\normalsize,
    at={(0,1.04)},anchor=south west},
  legend cell align=left,tick align=outside,clip=false
}}
\redcircuitexample{0}
\endgroup
    \end{minipage}
    \par
    \vspace{-5mm}
    \caption{Mixed CNOTs.}
    \label{fig:contention-loss-local-circuit}
  \end{subfigure}
  \hfill
  \begin{subfigure}[t]{0.33\textwidth}
    \centering
    \begin{minipage}[c][64mm][c]{\linewidth}
      \centering
      \begingroup
\usetikzlibrary{decorations.pathreplacing}
\newcommand{\redcircuitexample}[1]{%
  \begin{tikzpicture}[x=0.9mm,y=1.5mm,font=\normalsize]
    \path[use as bounding box] (-13,-25) rectangle (15,2);
    \foreach \i in {1,...,8}{
      \ifnum#1=0
        \ifnum\i<8
          \pgfmathsetmacro{\wy}{-3*(\i-1)}
          \pgfmathtruncatemacro{\qid}{\i}
        \else
          \def\wy{-23}
          \def\qid{17}
        \fi
      \else
        \ifnum\i<5
          \pgfmathsetmacro{\wy}{-3*(\i-1)}
          \pgfmathtruncatemacro{\qid}{\i}
        \else
          \pgfmathsetmacro{\wy}{-13-3*(\i-5)}
          \pgfmathtruncatemacro{\qid}{\i+12}
        \fi
      \fi
      \draw[black!55,line width=0.4pt] (0,\wy) -- (15,\wy);
      \node[anchor=east,inner sep=1pt] at (-1,\wy) {$Q_{\qid}$};
    }
    \ifnum#1=0
      \foreach \i in {1,2,3}{
        \pgfmathsetmacro{\cy}{-6*(\i-1)}
        \pgfmathsetmacro{\ty}{\cy-3}
        \draw[line width=0.55pt] (7.5,\cy) -- (7.5,\ty-1);
        \fill (7.5,\cy) circle (0.5mm);
        \draw[line width=0.55pt] (7.5,\ty) circle (0.8mm);
      }
      \draw[line width=0.55pt] (7.5,-18) -- (7.5,-24);
      \fill (7.5,-18) circle (0.5mm);
      \draw[line width=0.55pt] (7.5,-23) circle (0.8mm);
    \else
      \foreach \i in {1,...,4}{
        \pgfmathsetmacro{\gx}{2+3.5*(\i-1)}
        \pgfmathsetmacro{\cy}{-3*(\i-1)}
        \pgfmathsetmacro{\ty}{-13-3*(\i-1)}
        \draw[line width=0.55pt] (\gx,\cy) -- (\gx,\ty-1);
        \fill (\gx,\cy) circle (0.5mm);
        \draw[line width=0.55pt] (\gx,\ty) circle (0.8mm);
      }
    \fi
    \ifnum#1=0
      \draw[decorate,decoration={brace,amplitude=3pt},line width=0.45pt]
        (-7.5,-19.5) -- (-7.5,1.5);
      \node[rotate=90,inner sep=0pt] at (-12,-9) {ToR A};
      \draw[decorate,decoration={brace,amplitude=3pt},line width=0.45pt]
        (-7.5,-24.5) -- (-7.5,-21.5);
      \node[rotate=90,inner sep=0pt] at (-12,-23) {ToR B};
    \else
      \draw[decorate,decoration={brace,amplitude=3pt},line width=0.45pt]
        (-7.5,-10) -- (-7.5,1.5);
      \node[rotate=90,inner sep=0pt] at (-12,-4.25) {ToR A};
      \draw[decorate,decoration={brace,amplitude=3pt},line width=0.45pt]
        (-7.5,-23.5) -- (-7.5,-12);
      \node[rotate=90,inner sep=0pt] at (-12,-17.75) {ToR B};
    \fi
  \end{tikzpicture}%
}
\pgfplotsset{redtradeoff/.style={
  width=\linewidth,height=48mm,
  axis lines=left,xmin=2,xmax=8.3,ymin=0,ymax=1500,
  xtick={2,4,6,8},ytick={0,500,1000,1500},
  scaled y ticks=false,
  ylabel={Bell-pair completion time (ms)},
  tick label style={font=\normalsize},label style={font=\normalsize},
  legend style={draw=none,fill=none,font=\normalsize,
    at={(0,1.04)},anchor=south west},
  legend cell align=left,tick align=outside,clip=false
}}
\begin{tikzpicture}
\begin{axis}[redtradeoff,xmin=0,xmax=7.3,ymin=0,ymax=600,
  xtick={0,1,3,5,7},ytick={0,200,400,600},
  xlabel style={align=center},xlabel={Intra-rack CNOTs at A ($N$)\\One inter-rack CNOT}]
  \fill[black!10] (axis cs:0,260.977) -- (axis cs:1,288.119) -- (axis cs:2,315.261) -- (axis cs:3,342.403) -- (axis cs:4,369.545) -- (axis cs:5,396.687) -- (axis cs:6,423.829) -- (axis cs:7,450.971) -- (axis cs:7,260.977) -- (axis cs:6,260.977) -- (axis cs:5,260.977) -- (axis cs:4,260.977) -- (axis cs:3,260.977) -- (axis cs:2,260.977) -- (axis cs:1,260.977) -- (axis cs:0,260.977) -- cycle;
  \addplot[blue,thick] coordinates {(0,260.977) (1,288.119) (2,315.261) (3,342.403) (4,369.545) (5,396.687) (6,423.829) (7,450.971)};
  \addlegendentry{ToR A's BSM only}
  \addplot[red,thick,mark=*,mark size=1.3pt]
    coordinates {(0,260.977) (1,260.977) (2,260.977) (3,260.977) (4,260.977) (5,260.977) (6,260.977) (7,260.977)};
  \addlegendentry{ToR A or B's BSM}
  \node[align=center] (saving) at (axis cs:2.1,492)
    {Scheduling\\benefit};
  \draw[black!65,-{Latex[length=1.2mm]}]
    (saving.south) -- (axis cs:4.5,322.047);
\end{axis}
\end{tikzpicture}
\endgroup
    \end{minipage}
    \par
    \vspace{-5mm}
    \caption{Scheduling benefit.}
    \label{fig:contention-loss-scheduling}
  \end{subfigure}
  \hfill
  \begin{subfigure}[t]{0.15\textwidth}
    \centering
    \begin{minipage}[c][64mm][c]{\linewidth}
      \centering
      \begingroup
\usetikzlibrary{decorations.pathreplacing}
\newcommand{\redcircuitexample}[1]{%
  \begin{tikzpicture}[x=0.9mm,y=1.5mm,font=\normalsize]
    \path[use as bounding box] (-13,-25) rectangle (15,2);
    \foreach \i in {1,...,8}{
      \ifnum#1=0
        \ifnum\i<8
          \pgfmathsetmacro{\wy}{-3*(\i-1)}
          \pgfmathtruncatemacro{\qid}{\i}
        \else
          \def\wy{-23}
          \def\qid{17}
        \fi
      \else
        \ifnum\i<5
          \pgfmathsetmacro{\wy}{-3*(\i-1)}
          \pgfmathtruncatemacro{\qid}{\i}
        \else
          \pgfmathsetmacro{\wy}{-13-3*(\i-5)}
          \pgfmathtruncatemacro{\qid}{\i+12}
        \fi
      \fi
      \draw[black!55,line width=0.4pt] (0,\wy) -- (15,\wy);
      \node[anchor=east,inner sep=1pt] at (-1,\wy) {$Q_{\qid}$};
    }
    \ifnum#1=0
      \foreach \i in {1,2,3}{
        \pgfmathsetmacro{\cy}{-6*(\i-1)}
        \pgfmathsetmacro{\ty}{\cy-3}
        \draw[line width=0.55pt] (7.5,\cy) -- (7.5,\ty-1);
        \fill (7.5,\cy) circle (0.5mm);
        \draw[line width=0.55pt] (7.5,\ty) circle (0.8mm);
      }
      \draw[line width=0.55pt] (7.5,-18) -- (7.5,-24);
      \fill (7.5,-18) circle (0.5mm);
      \draw[line width=0.55pt] (7.5,-23) circle (0.8mm);
    \else
      \foreach \i in {1,...,4}{
        \pgfmathsetmacro{\gx}{2+3.5*(\i-1)}
        \pgfmathsetmacro{\cy}{-3*(\i-1)}
        \pgfmathsetmacro{\ty}{-13-3*(\i-1)}
        \draw[line width=0.55pt] (\gx,\cy) -- (\gx,\ty-1);
        \fill (\gx,\cy) circle (0.5mm);
        \draw[line width=0.55pt] (\gx,\ty) circle (0.8mm);
      }
    \fi
    \ifnum#1=0
      \draw[decorate,decoration={brace,amplitude=3pt},line width=0.45pt]
        (-7.5,-19.5) -- (-7.5,1.5);
      \node[rotate=90,inner sep=0pt] at (-12,-9) {ToR A};
      \draw[decorate,decoration={brace,amplitude=3pt},line width=0.45pt]
        (-7.5,-24.5) -- (-7.5,-21.5);
      \node[rotate=90,inner sep=0pt] at (-12,-23) {ToR B};
    \else
      \draw[decorate,decoration={brace,amplitude=3pt},line width=0.45pt]
        (-7.5,-10) -- (-7.5,1.5);
      \node[rotate=90,inner sep=0pt] at (-12,-4.25) {ToR A};
      \draw[decorate,decoration={brace,amplitude=3pt},line width=0.45pt]
        (-7.5,-23.5) -- (-7.5,-12);
      \node[rotate=90,inner sep=0pt] at (-12,-17.75) {ToR B};
    \fi
  \end{tikzpicture}%
}
\pgfplotsset{redtradeoff/.style={
  width=\linewidth,height=48mm,
  axis lines=left,xmin=2,xmax=8.3,ymin=0,ymax=1500,
  xtick={2,4,6,8},ytick={0,500,1000,1500},
  scaled y ticks=false,
  ylabel={Bell-pair completion time (ms)},
  tick label style={font=\normalsize},label style={font=\normalsize},
  legend style={draw=none,fill=none,font=\normalsize,
    at={(0,1.04)},anchor=south west},
  legend cell align=left,tick align=outside,clip=false
}}
\redcircuitexample{1}
\endgroup
    \end{minipage}
    \par
    \vspace{-5mm}
    \caption{Inter-rack CNOTs.}
    \label{fig:contention-loss-cross-circuit}
  \end{subfigure}
  \hfill
  \begin{subfigure}[t]{0.33\textwidth}
    \centering
    \begin{minipage}[c][64mm][c]{\linewidth}
      \centering
      \begingroup
\usetikzlibrary{decorations.pathreplacing}
\newcommand{\redcircuitexample}[1]{%
  \begin{tikzpicture}[x=0.9mm,y=1.5mm,font=\normalsize]
    \path[use as bounding box] (-13,-25) rectangle (15,2);
    \foreach \i in {1,...,8}{
      \ifnum#1=0
        \ifnum\i<8
          \pgfmathsetmacro{\wy}{-3*(\i-1)}
          \pgfmathtruncatemacro{\qid}{\i}
        \else
          \def\wy{-23}
          \def\qid{17}
        \fi
      \else
        \ifnum\i<5
          \pgfmathsetmacro{\wy}{-3*(\i-1)}
          \pgfmathtruncatemacro{\qid}{\i}
        \else
          \pgfmathsetmacro{\wy}{-13-3*(\i-5)}
          \pgfmathtruncatemacro{\qid}{\i+12}
        \fi
      \fi
      \draw[black!55,line width=0.4pt] (0,\wy) -- (15,\wy);
      \node[anchor=east,inner sep=1pt] at (-1,\wy) {$Q_{\qid}$};
    }
    \ifnum#1=0
      \foreach \i in {1,2,3}{
        \pgfmathsetmacro{\cy}{-6*(\i-1)}
        \pgfmathsetmacro{\ty}{\cy-3}
        \draw[line width=0.55pt] (7.5,\cy) -- (7.5,\ty-1);
        \fill (7.5,\cy) circle (0.5mm);
        \draw[line width=0.55pt] (7.5,\ty) circle (0.8mm);
      }
      \draw[line width=0.55pt] (7.5,-18) -- (7.5,-24);
      \fill (7.5,-18) circle (0.5mm);
      \draw[line width=0.55pt] (7.5,-23) circle (0.8mm);
    \else
      \foreach \i in {1,...,4}{
        \pgfmathsetmacro{\gx}{2+3.5*(\i-1)}
        \pgfmathsetmacro{\cy}{-3*(\i-1)}
        \pgfmathsetmacro{\ty}{-13-3*(\i-1)}
        \draw[line width=0.55pt] (\gx,\cy) -- (\gx,\ty-1);
        \fill (\gx,\cy) circle (0.5mm);
        \draw[line width=0.55pt] (\gx,\ty) circle (0.8mm);
      }
    \fi
    \ifnum#1=0
      \draw[decorate,decoration={brace,amplitude=3pt},line width=0.45pt]
        (-7.5,-19.5) -- (-7.5,1.5);
      \node[rotate=90,inner sep=0pt] at (-12,-9) {ToR A};
      \draw[decorate,decoration={brace,amplitude=3pt},line width=0.45pt]
        (-7.5,-24.5) -- (-7.5,-21.5);
      \node[rotate=90,inner sep=0pt] at (-12,-23) {ToR B};
    \else
      \draw[decorate,decoration={brace,amplitude=3pt},line width=0.45pt]
        (-7.5,-10) -- (-7.5,1.5);
      \node[rotate=90,inner sep=0pt] at (-12,-4.25) {ToR A};
      \draw[decorate,decoration={brace,amplitude=3pt},line width=0.45pt]
        (-7.5,-23.5) -- (-7.5,-12);
      \node[rotate=90,inner sep=0pt] at (-12,-17.75) {ToR B};
    \fi
  \end{tikzpicture}%
}
\pgfplotsset{redtradeoff/.style={
  width=\linewidth,height=48mm,
  axis lines=left,xmin=2,xmax=8.3,ymin=0,ymax=1500,
  xtick={2,4,6,8},ytick={0,500,1000,1500},
  scaled y ticks=false,
  ylabel={Bell-pair completion time (ms)},
  tick label style={font=\normalsize},label style={font=\normalsize},
  legend style={draw=none,fill=none,font=\normalsize,
    at={(0,1.04)},anchor=south west},
  legend cell align=left,tick align=outside,clip=false
}}
\begin{tikzpicture}
\begin{axis}[redtradeoff,legend style={at={(1,1.04)},anchor=south east},
  xlabel style={align=center},xlabel={Remote CNOTs ($N$)\\$N$ participating QPUs per ToR}]
  \fill[black!10] (axis cs:2,260.977) -- (axis cs:8,1043.909)
    -- (axis cs:8,108.568) -- (axis cs:2,27.142) -- cycle;
  \addplot[orange!80!black,thick,mark=*,mark size=1.3pt]
    coordinates {(2,260.977) (4,521.954) (6,782.932) (8,1043.909)};
  \addlegendentry{With wavelength conversion}
  \addplot[black,dashed,thick] coordinates {(2,27.142) (4,54.284) (6,81.426) (8,108.568)};
  \addlegendentry{Without wavelength conversion}
  \node[align=center] (penalty) at (axis cs:4,1080)
    {Fixed-connectivity\\path penalty};
  \draw[black!65,-{Latex[length=1.2mm]}]
    (penalty.south) -- (axis cs:5.4,388.961);
  \node[anchor=south east,inner sep=2pt] at (axis cs:8,1043.9) {1043.9};
  \node[anchor=south east,inner sep=2pt] at (axis cs:8,108.6) {108.6};
\end{axis}
\end{tikzpicture}
\endgroup
    \end{minipage}
    \par
    \vspace{-5mm}
    \caption{Fixed-connectivity path penalty.}
    \label{fig:contention-loss-placement}
  \end{subfigure}
  \vspace{-3mm}
  \caption{Scheduling reduces BSM contention, but inter-rack pairs still incur conversion loss. \textit{(a, b)} We vary the number of intra-rack CNOTs at A, $N$, while retaining one A-B CNOT. Generating the A-B pair at B lets A serve its local pairs concurrently without additional conversion. For the $N=3$ circuit shown in (a), completion falls from 342.4 to 261.0~ms, a 23.8\% reduction compared to serving all four pairs at A. \textit{(c, d)} When every pair spans the two racks, balancing BSM use still leaves one photon per pair traversing a converted path. The shaded gap relative to equally balanced native local generation shows the path penalty that scheduling cannot remove. Each ToR has 16 QPUs and one BSM (we only show participating QPUs).}
  \vspace{-3mm}
  \label{fig:contention-loss}
\end{figure*}
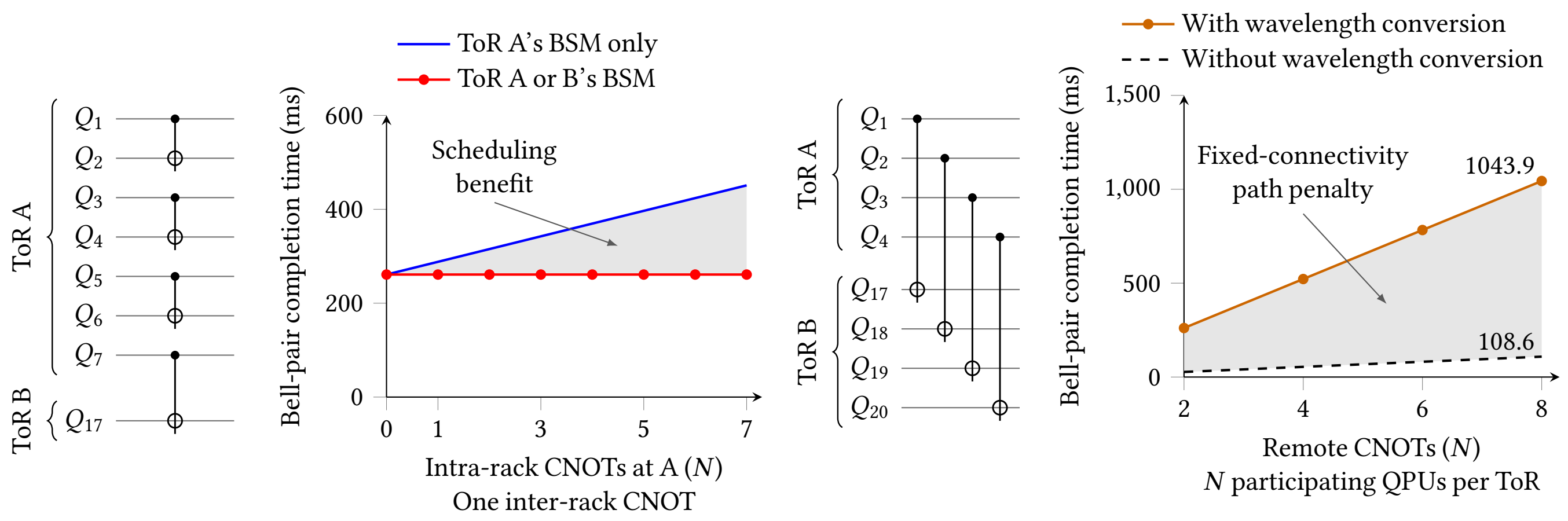

In switch-based Clos architectures with BSMs and QFCs at the ToRs~\cite{zhang2025switchqnet,shapourian2025quantum}, fixed connections determine which BSM pool each QPU can reach without wavelength conversion. Scheduling can relieve contention by distributing Bell-pair generation across pools, but cannot change the photon paths used to reach them.

To illustrate this tradeoff, we consider an example with two racks with 16 QPUs and one BSM each, using two circuits (Figure~\ref{fig:contention-loss}) that highlight how BSM contention and photon-path loss affect communication completion time. QPUs $Q_1$ through $Q_{16}$ connect to ToR A, and $Q_{17}$ through $Q_{32}$ to ToR B. Two aggregation switches connect the ToRs through QFC-equipped uplinks, with one optical channel per link, allowing both BSMs to serve inter-rack pairs concurrently. Local generation avoids conversion, while reaching the other ToR's BSM requires two QFCs and an aggregation switch. The CNOTs are ready simultaneously, connect disjoint QPU pairs, and require one Bell pair each. We estimate the time to establish all pairs using expected generation times and 10~$\mu s$ of path setup per pair, reserving BSMs and links through setup and retries. Our evaluation (\S\ref{sec:evaluation}) models stochastic attempts and retries.

For Figures~\ref{fig:contention-loss} and~\ref{fig:sharing-examples}, we use 2.4~dB loss per QFC~\cite{Li_2016, vanleent2022entangling} and 0.3~dB per $2\times 2$ switching element. A Bene\v{s} path traverses nine elements at each 32-port ToR and one at each two-port aggregation switch, contributing 2.7 and 0.3~dB, respectively. Each fiber link is 50~m with attenuation of 0.2~dB/km. We set effective photon efficiency based on $^{88}\mathrm{Sr}^{+}$ trapped-ion hardware~\cite{stephenson2020high}. BSM acceptance probability is 0.5, and we use an effective attempt cycle of 1.198~$\mu s$ plus round-trip propagation. Including path setup, a pair takes 27.1~ms locally and 261.0~ms when one photon crosses ToRs.

\myitem{Scheduling relieves BSM contention:} We combine $N$ intra-rack CNOTs at A with one A-B CNOT, varying $N$ from zero to seven and selecting the BSM assignment with the shortest completion time (Figure~\ref{fig:contention-loss-scheduling}). For the $N=3$ circuit in Figure~\ref{fig:contention-loss-local-circuit}, serving all four pairs at A takes 342.4~ms. Assigning the inter-rack pair to B lets A serve its local pairs concurrently, reducing completion time to 261.0~ms while preserving the same photon-path loss.

\myitem{Fixed connectivity imposes photon-path loss:} In Figures~\ref{fig:contention-loss-cross-circuit} and~\ref{fig:contention-loss-placement}, every CNOT connects a QPU at A to a QPU at B. Either BSM can generate each pair, but one photon must cross ToRs regardless of the choice. Balancing generation across both BSMs minimizes completion time, yet eight pairs still take 1043.9~ms, compared to 108.6~ms through native-wavelength paths with the same BSM capacity. The shaded gap shows the additional completion time imposed by fixed photon paths even after scheduling balances BSM use.

As communication changes across circuit phases, the QPUs that benefit from sharing a pool can change as well. Fixed connectivity locks programs into a tradeoff between local BSM contention and photon-path loss: scheduling can redistribute generation, but cannot change which QPUs reach each pool without wavelength conversion.

\smallskip
\takeaway{Scheduling can balance Bell-pair generation across BSMs, but fixed QPU-to-ToR connectivity determines which pairs require wavelength conversion. Removing that loss requires changing which QPUs share native-wavelength access to a pool.}

\subsection{Towards Reconfigurable BSM Sharing}
\label{sec:reconfigurable-sharing}

We next ask whether changing which QPUs share a BSM pool can overcome these limits. We use the same circuits and 32 QPUs, with one BSM and 16 QPUs per pool, but allow the assignments to change. Pairs within a pool avoid conversion, while pairs spanning pools still require it. Reconfigurable paths add an aggregate loss of 0.5~dB per photon, increasing local service time from 27.1 to 34.2~ms, including path setup. We first compare groupings within a phase, then consider when the savings justify changing them.

\begin{figure}[t]
  \begingroup
  \centering
  \input{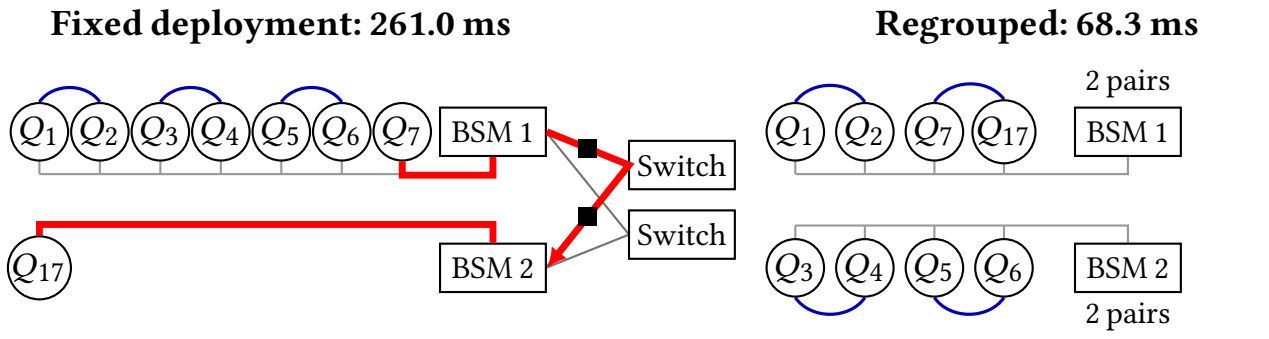}
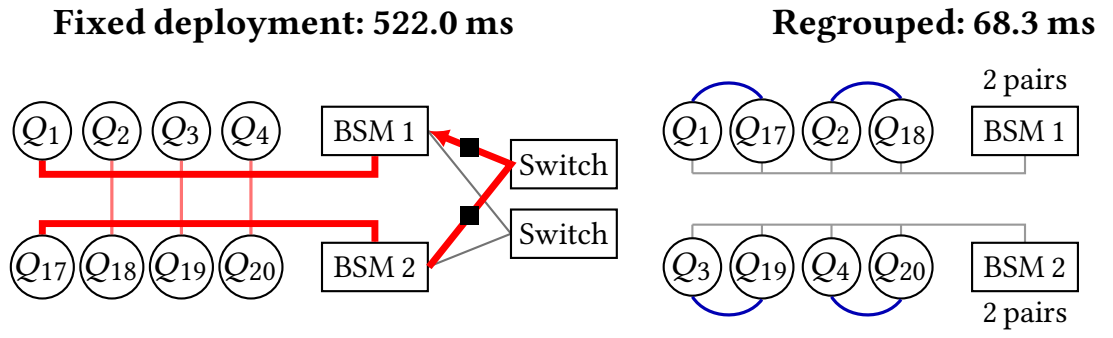
  \begin{subfigure}[b]{\columnwidth}
    \centering
    \begin{tikzpicture}[x=1mm,y=1mm]
\begin{scope}[xshift=0mm]
\node[title] at (16,7) {Fixed deployment: 261.0~ms};
\draw[access] (0,0) -- (0,-2.8);
\draw[access] (4,0) -- (4,-2.8);
\draw[access] (8,0) -- (8,-2.8);
\draw[access] (12,0) -- (12,-2.8);
\draw[access] (16,0) -- (16,-2.8);
\draw[access] (20,0) -- (20,-2.8);
\draw[access] (24,0) -- (24,-2.8);
\draw[access] (0,-2.8) -| (30,0);
\draw[access] (0,-9) -- (0,-6.2);
\draw[access] (0,-6.2) -| (30,-9);
\draw[link] (33.5,0) -- (39,-2.2);
\draw[link] (33.5,-9) -- (39,-2.2);
\draw[link] (33.5,0) -- (39,-6.8);
\draw[link] (33.5,-9) -- (39,-6.8);
\node[spine] at (42.5,-2.2) {Switch};
\node[spine] at (42.5,-6.8) {Switch};
\node[qpu] (q1) at (0,0) {$Q_{1}$};
\node[qpu] (q2) at (4,0) {$Q_{2}$};
\node[qpu] (q3) at (8,0) {$Q_{3}$};
\node[qpu] (q4) at (12,0) {$Q_{4}$};
\node[qpu] (q5) at (16,0) {$Q_{5}$};
\node[qpu] (q6) at (20,0) {$Q_{6}$};
\node[qpu] (q7) at (24,0) {$Q_{7}$};
\node[qpu] (q17) at (0,-9) {$Q_{17}$};
\node[tor] (t0) at (30,0) {BSM 1};
\node[tor] (t1) at (30,-9) {BSM 2};
\draw[lpair] (q1.north) to[bend left=55] (q2.north);
\draw[lpair] (q3.north) to[bend left=55] (q4.north);
\draw[lpair] (q5.north) to[bend left=55] (q6.north);
\draw[route,crosspair] (q7.south) |- ($(t0.south)+(0,-1.2)$) -- (t0.south);
\draw[route,crosspair] (q17.north) |- ($(t1.north)+(0,1.2)$) -- (t1.north);
\draw[arrive,crosspair] (t0.east) -- (39,-2.2) -- (t1.east);
\node[qfc] at (36.25,-1.1) {};
\node[qfc] at (36.25,-5.6) {};
\end{scope}
\begin{scope}[xshift=50mm]
\node[title] at (16,7) {Regrouped: 68.3~ms};
\draw[access] (0.0,0) -- (0.0,-2.8);
\draw[access] (4.6,0) -- (4.6,-2.8);
\draw[access] (9.2,0) -- (9.2,-2.8);
\draw[access] (13.8,0) -- (13.8,-2.8);
\draw[access] (0,-2.8) -| (22,0);
\draw[access] (0.0,-9) -- (0.0,-6.2);
\draw[access] (4.6,-9) -- (4.6,-6.2);
\draw[access] (9.2,-9) -- (9.2,-6.2);
\draw[access] (13.8,-9) -- (13.8,-6.2);
\draw[access] (0,-6.2) -| (22,-9);
\node[qpu] (q1) at (0.0,0) {$Q_{1}$};
\node[qpu] (q2) at (4.6,0) {$Q_{2}$};
\node[qpu] (q7) at (9.2,0) {$Q_{7}$};
\node[qpu] (q17) at (13.8,0) {$Q_{17}$};
\node[qpu] (q3) at (0.0,-9) {$Q_{3}$};
\node[qpu] (q4) at (4.6,-9) {$Q_{4}$};
\node[qpu] (q5) at (9.2,-9) {$Q_{5}$};
\node[qpu] (q6) at (13.8,-9) {$Q_{6}$};
\node[tor] (t0) at (22,0) {BSM 1};
\node[load,anchor=south] at (22,2) {2 pairs};
\node[tor] (t1) at (22,-9) {BSM 2};
\node[load,anchor=north] at (22,-11) {2 pairs};
\draw[lpair] (q1.north) to[bend left=55] (q2.north);
\draw[lpair] (q3.south) to[bend right=55] (q4.south);
\draw[lpair] (q5.south) to[bend right=55] (q6.south);
\draw[lpair] (q7.north) to[bend left=55] (q17.north);
\end{scope}
\end{tikzpicture}
    \vspace{-2mm}
    \caption{Balanced BSM use for the mixed CNOTs.}
    \label{fig:motivation-ex2}\label{fig:sharing-example-contention}
  \end{subfigure}

  \medskip
    \begin{subfigure}[b]{\columnwidth}
    \centering
    \begin{tikzpicture}[x=1mm,y=1mm]
\begin{scope}[xshift=0mm]
\node[title] at (16,7) {Fixed deployment: 522.0~ms};
\draw[access] (0.0,0) -- (0.0,-2.8);
\draw[access] (4.6,0) -- (4.6,-2.8);
\draw[access] (9.2,0) -- (9.2,-2.8);
\draw[access] (13.8,0) -- (13.8,-2.8);
\draw[access] (0,-2.8) -| (22,0);
\draw[access] (0.0,-9) -- (0.0,-6.2);
\draw[access] (4.6,-9) -- (4.6,-6.2);
\draw[access] (9.2,-9) -- (9.2,-6.2);
\draw[access] (13.8,-9) -- (13.8,-6.2);
\draw[access] (0,-6.2) -| (22,-9);
\draw[link] (25.5,0) -- (31,-2.2);
\draw[link] (25.5,-9) -- (31,-2.2);
\draw[link] (25.5,0) -- (31,-6.8);
\draw[link] (25.5,-9) -- (31,-6.8);
\node[spine] at (34.5,-2.2) {Switch};
\node[spine] at (34.5,-6.8) {Switch};
\node[qpu] (q1) at (0.0,0) {$Q_{1}$};
\node[qpu] (q2) at (4.6,0) {$Q_{2}$};
\node[qpu] (q3) at (9.2,0) {$Q_{3}$};
\node[qpu] (q4) at (13.8,0) {$Q_{4}$};
\node[qpu] (q17) at (0.0,-9) {$Q_{17}$};
\node[qpu] (q18) at (4.6,-9) {$Q_{18}$};
\node[qpu] (q19) at (9.2,-9) {$Q_{19}$};
\node[qpu] (q20) at (13.8,-9) {$Q_{20}$};
\node[tor] (t0) at (22,0) {BSM 1};
\node[tor] (t1) at (22,-9) {BSM 2};
\draw[xpair,crosspair!55] (q2) -- (q18);
\draw[xpair,crosspair!55] (q3) -- (q19);
\draw[xpair,crosspair!55] (q4) -- (q20);
\draw[route,crosspair] (q1.south) |- ($(t0.south)+(0,-1.2)$) -- (t0.south);
\draw[route,crosspair] (q17.north) |- ($(t1.north)+(0,1.2)$) -- (t1.north);
\draw[arrive,crosspair] (t1.east) -- (31,-2.2) -- (t0.east);
\node[qfc] at (28.25,-1.1) {};
\node[qfc] at (28.25,-5.6) {};
\end{scope}
\begin{scope}[xshift=43mm]
\node[title] at (16,7) {Regrouped: 68.3~ms};
\draw[access] (0.0,0) -- (0.0,-2.8);
\draw[access] (4.6,0) -- (4.6,-2.8);
\draw[access] (9.2,0) -- (9.2,-2.8);
\draw[access] (13.8,0) -- (13.8,-2.8);
\draw[access] (0,-2.8) -| (22,0);
\draw[access] (0.0,-9) -- (0.0,-6.2);
\draw[access] (4.6,-9) -- (4.6,-6.2);
\draw[access] (9.2,-9) -- (9.2,-6.2);
\draw[access] (13.8,-9) -- (13.8,-6.2);
\draw[access] (0,-6.2) -| (22,-9);
\node[qpu] (q1) at (0.0,0) {$Q_{1}$};
\node[qpu] (q17) at (4.6,0) {$Q_{17}$};
\node[qpu] (q2) at (9.2,0) {$Q_{2}$};
\node[qpu] (q18) at (13.8,0) {$Q_{18}$};
\node[qpu] (q3) at (0.0,-9) {$Q_{3}$};
\node[qpu] (q19) at (4.6,-9) {$Q_{19}$};
\node[qpu] (q4) at (9.2,-9) {$Q_{4}$};
\node[qpu] (q20) at (13.8,-9) {$Q_{20}$};
\node[tor] (t0) at (22,0) {BSM 1};
\node[load,anchor=south] at (22,2) {2 pairs};
\node[tor] (t1) at (22,-9) {BSM 2};
\node[load,anchor=north] at (22,-11) {2 pairs};
\draw[lpair] (q1.north) to[bend left=55] (q17.north);
\draw[lpair] (q2.north) to[bend left=55] (q18.north);
\draw[lpair] (q3.south) to[bend right=55] (q19.south);
\draw[lpair] (q4.south) to[bend right=55] (q20.south);
\end{scope}
\end{tikzpicture}
    \vspace{-2mm}
    \caption{Shorter photon paths for the inter-rack CNOTs.}
    \label{fig:motivation-ex1}\label{fig:sharing-example-locality}
  \end{subfigure}

  \endgroup
  \vspace{-2mm}
  \caption{Regrouping changes photon paths and BSM contention for the circuits in Figure~\ref{fig:contention-loss}. \textit{(a)} Regrouping the mixed CNOTs avoids conversion while balancing generation across both BSMs. \textit{(b)} Bringing each inter-rack pair to the same pool avoids conversion.} 
  \label{fig:sharing-examples}
  \vspace{-5mm}
\end{figure}

\myitem{Regrouping removes conversion imposed by fixed connectivity:} For the four inter-rack CNOTs from Figure~\ref{fig:contention-loss-cross-circuit}, scheduling balances BSM use, but every pair requires conversion. Regrouping assigns both QPUs of each pair to the same pool, reducing completion time from 522.0 to 68.3~ms despite the additional path loss (Figure~\ref{fig:sharing-example-locality}). With each BSM still serving two pairs, the improvement comes from changing the photon paths used to reach them.

\myitem{Grouping must balance BSM use as well as locality:} For the mixed circuit from Figure~\ref{fig:contention-loss-local-circuit}, scheduling serves the local pairs at A and the inter-rack pair at B, whose converted path determines completion time. Regrouping gives each pool two local pairs, reducing completion time from 261.0 to 68.3~ms (Figure~\ref{fig:sharing-example-contention}). Placing all four pairs in one pool would also avoid conversion, but force them to share a single BSM. Choosing between these groupings requires considering contention as well as locality.

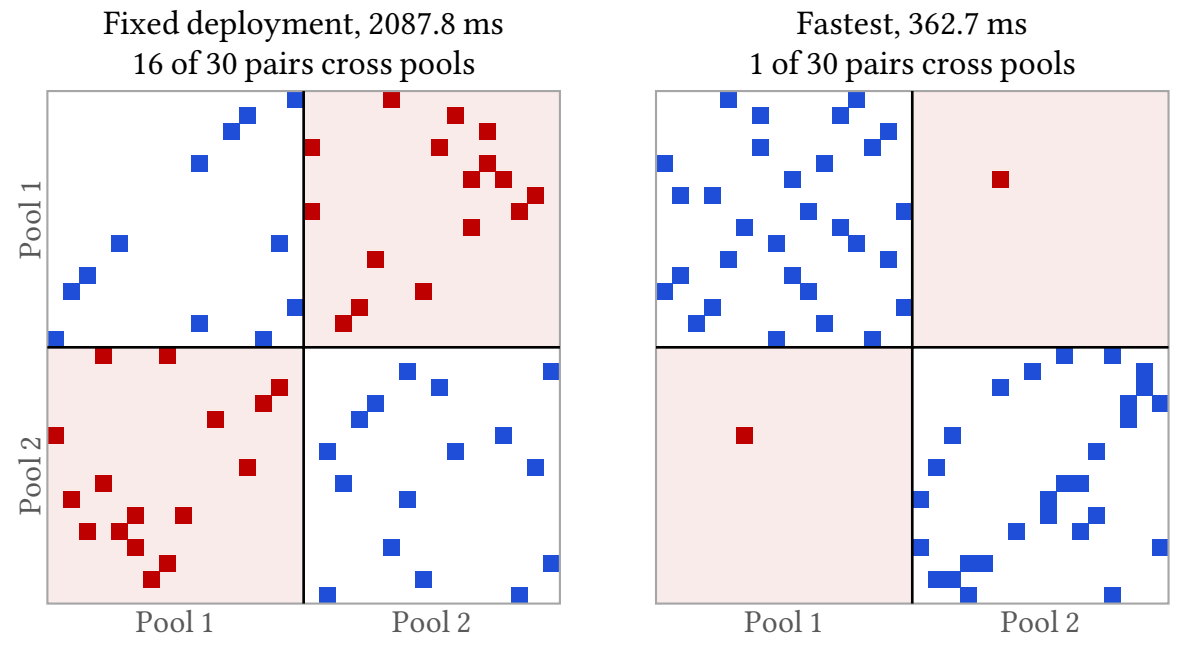
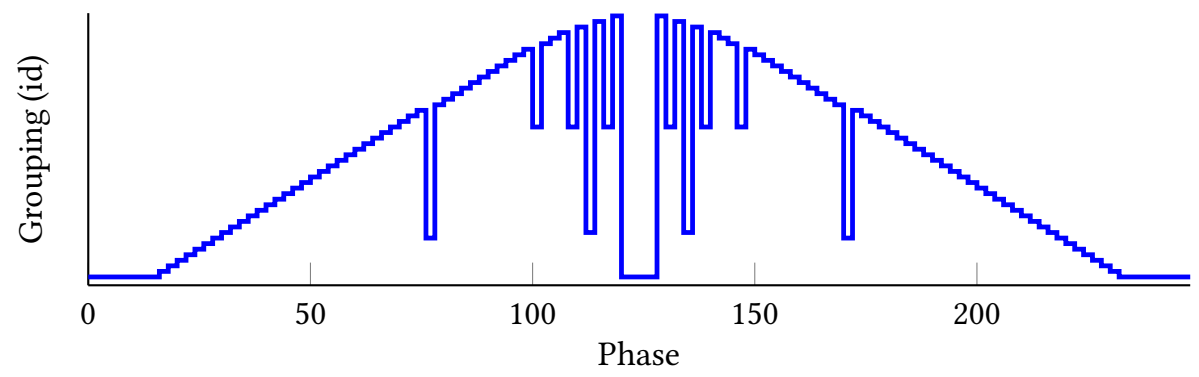
\begin{figure}[t]
  \centering
  \begin{subfigure}[b]{\columnwidth}
    \centering
    \definecolor{phaselocal}{HTML}{1F4FD9}
\colorlet{phasecross}{myred}
\begin{tikzpicture}[
    local/.style={fill=phaselocal},
    cross/.style={fill=phasecross},
    crosszone/.style={fill=phasecross!8},
    frame/.style={draw=black!35, line width=0.4pt},
    pool/.style={draw=black, line width=0.5pt},
    figtext/.style={font=\fontsize{6.5}{7.5}\selectfont, align=center, inner sep=1.5pt},
    tick/.style={font=\fontsize{6}{7}\selectfont, text=black!65, inner sep=1pt}]
  \begin{scope}[x={0.40\columnwidth/32}, y={0.40\columnwidth/32}]
  \begin{scope}[shift={(0,0)}]
    \fill[crosszone] (16,0) rectangle (32,-16);
    \fill[crosszone] (0,-16) rectangle (16,-32);
    \fill[cross] (16,-7) rectangle ++(1,-1);
    \fill[cross] (7,-16) rectangle ++(1,-1);
    \fill[cross] (29,-7) rectangle ++(1,-1);
    \fill[cross] (7,-29) rectangle ++(1,-1);
    \fill[cross] (20,-10) rectangle ++(1,-1);
    \fill[cross] (10,-20) rectangle ++(1,-1);
    \fill[cross] (16,-3) rectangle ++(1,-1);
    \fill[cross] (3,-16) rectangle ++(1,-1);
    \fill[local] (31,-29) rectangle ++(1,-1);
    \fill[local] (29,-31) rectangle ++(1,-1);
    \fill[local] (20,-19) rectangle ++(1,-1);
    \fill[local] (19,-20) rectangle ++(1,-1);
    \fill[cross] (24,-3) rectangle ++(1,-1);
    \fill[cross] (3,-24) rectangle ++(1,-1);
    \fill[local] (31,-17) rectangle ++(1,-1);
    \fill[local] (17,-31) rectangle ++(1,-1);
    \fill[cross] (19,-13) rectangle ++(1,-1);
    \fill[cross] (13,-19) rectangle ++(1,-1);
    \fill[local] (24,-18) rectangle ++(1,-1);
    \fill[local] (18,-24) rectangle ++(1,-1);
    \fill[local] (22,-17) rectangle ++(1,-1);
    \fill[local] (17,-22) rectangle ++(1,-1);
    \fill[local] (15,-13) rectangle ++(1,-1);
    \fill[local] (13,-15) rectangle ++(1,-1);
    \fill[cross] (18,-14) rectangle ++(1,-1);
    \fill[cross] (14,-18) rectangle ++(1,-1);
    \fill[local] (25,-22) rectangle ++(1,-1);
    \fill[local] (22,-25) rectangle ++(1,-1);
    \fill[local] (15,0) rectangle ++(1,-1);
    \fill[local] (0,-15) rectangle ++(1,-1);
    \fill[local] (14,-9) rectangle ++(1,-1);
    \fill[local] (9,-14) rectangle ++(1,-1);
    \fill[cross] (25,-1) rectangle ++(1,-1);
    \fill[cross] (1,-25) rectangle ++(1,-1);
    \fill[cross] (21,0) rectangle ++(1,-1);
    \fill[cross] (0,-21) rectangle ++(1,-1);
    \fill[local] (9,-4) rectangle ++(1,-1);
    \fill[local] (4,-9) rectangle ++(1,-1);
    \fill[local] (12,-1) rectangle ++(1,-1);
    \fill[local] (1,-12) rectangle ++(1,-1);
    \fill[local] (28,-21) rectangle ++(1,-1);
    \fill[local] (21,-28) rectangle ++(1,-1);
    \fill[cross] (27,-4) rectangle ++(1,-1);
    \fill[cross] (4,-27) rectangle ++(1,-1);
    \fill[cross] (23,-12) rectangle ++(1,-1);
    \fill[cross] (12,-23) rectangle ++(1,-1);
    \fill[cross] (28,-5) rectangle ++(1,-1);
    \fill[cross] (5,-28) rectangle ++(1,-1);
    \fill[cross] (27,-2) rectangle ++(1,-1);
    \fill[cross] (2,-27) rectangle ++(1,-1);
    \fill[local] (30,-23) rectangle ++(1,-1);
    \fill[local] (23,-30) rectangle ++(1,-1);
    \fill[cross] (26,-5) rectangle ++(1,-1);
    \fill[cross] (5,-26) rectangle ++(1,-1);
    \fill[local] (11,-2) rectangle ++(1,-1);
    \fill[local] (2,-11) rectangle ++(1,-1);
    \fill[cross] (30,-6) rectangle ++(1,-1);
    \fill[cross] (6,-30) rectangle ++(1,-1);
    \fill[cross] (26,-8) rectangle ++(1,-1);
    \fill[cross] (8,-26) rectangle ++(1,-1);
    \draw[frame] (0,0) rectangle (32,-32);
    \draw[pool] (16,0) -- (16,-32) (0,-16) -- (32,-16);
    \node[figtext, above] at (16.0,0) {Fixed deployment, 2087.8~ms\\16 of 30 pairs cross pools};
    \node[tick, below] at (8.0,-32) {Pool 1};
    \node[tick, below] at (24.0,-32) {Pool 2};
    \node[tick, rotate=90, above] at (0,-8.0) {Pool 1};
    \node[tick, rotate=90, above] at (0,-24.0) {Pool 2};
  \end{scope}
  \begin{scope}[shift={(38,0)}]
    \fill[crosszone] (16,0) rectangle (32,-16);
    \fill[crosszone] (0,-16) rectangle (16,-32);
    \fill[local] (6,-3) rectangle ++(1,-1);
    \fill[local] (3,-6) rectangle ++(1,-1);
    \fill[local] (13,-3) rectangle ++(1,-1);
    \fill[local] (3,-13) rectangle ++(1,-1);
    \fill[local] (27,-22) rectangle ++(1,-1);
    \fill[local] (22,-27) rectangle ++(1,-1);
    \fill[local] (6,-1) rectangle ++(1,-1);
    \fill[local] (1,-6) rectangle ++(1,-1);
    \fill[local] (15,-13) rectangle ++(1,-1);
    \fill[local] (13,-15) rectangle ++(1,-1);
    \fill[local] (27,-26) rectangle ++(1,-1);
    \fill[local] (26,-27) rectangle ++(1,-1);
    \fill[local] (11,-1) rectangle ++(1,-1);
    \fill[local] (1,-11) rectangle ++(1,-1);
    \fill[local] (15,-7) rectangle ++(1,-1);
    \fill[local] (7,-15) rectangle ++(1,-1);
    \fill[local] (26,-24) rectangle ++(1,-1);
    \fill[local] (24,-26) rectangle ++(1,-1);
    \fill[local] (11,-8) rectangle ++(1,-1);
    \fill[local] (8,-11) rectangle ++(1,-1);
    \fill[local] (9,-7) rectangle ++(1,-1);
    \fill[local] (7,-9) rectangle ++(1,-1);
    \fill[local] (25,-24) rectangle ++(1,-1);
    \fill[local] (24,-25) rectangle ++(1,-1);
    \fill[local] (8,-5) rectangle ++(1,-1);
    \fill[local] (5,-8) rectangle ++(1,-1);
    \fill[local] (12,-9) rectangle ++(1,-1);
    \fill[local] (9,-12) rectangle ++(1,-1);
    \fill[local] (25,-16) rectangle ++(1,-1);
    \fill[local] (16,-25) rectangle ++(1,-1);
    \fill[cross] (5,-21) rectangle ++(1,-1);
    \fill[cross] (21,-5) rectangle ++(1,-1);
    \fill[local] (12,0) rectangle ++(1,-1);
    \fill[local] (0,-12) rectangle ++(1,-1);
    \fill[local] (28,-16) rectangle ++(1,-1);
    \fill[local] (16,-28) rectangle ++(1,-1);
    \fill[local] (21,-18) rectangle ++(1,-1);
    \fill[local] (18,-21) rectangle ++(1,-1);
    \fill[local] (4,0) rectangle ++(1,-1);
    \fill[local] (0,-4) rectangle ++(1,-1);
    \fill[local] (31,-28) rectangle ++(1,-1);
    \fill[local] (28,-31) rectangle ++(1,-1);
    \fill[local] (30,-18) rectangle ++(1,-1);
    \fill[local] (18,-30) rectangle ++(1,-1);
    \fill[local] (10,-4) rectangle ++(1,-1);
    \fill[local] (4,-10) rectangle ++(1,-1);
    \fill[local] (31,-19) rectangle ++(1,-1);
    \fill[local] (19,-31) rectangle ++(1,-1);
    \fill[local] (30,-17) rectangle ++(1,-1);
    \fill[local] (17,-30) rectangle ++(1,-1);
    \fill[local] (14,-10) rectangle ++(1,-1);
    \fill[local] (10,-14) rectangle ++(1,-1);
    \fill[local] (29,-19) rectangle ++(1,-1);
    \fill[local] (19,-29) rectangle ++(1,-1);
    \fill[local] (23,-17) rectangle ++(1,-1);
    \fill[local] (17,-23) rectangle ++(1,-1);
    \fill[local] (14,-2) rectangle ++(1,-1);
    \fill[local] (2,-14) rectangle ++(1,-1);
    \fill[local] (29,-20) rectangle ++(1,-1);
    \fill[local] (20,-29) rectangle ++(1,-1);
    \draw[frame] (0,0) rectangle (32,-32);
    \draw[pool] (16,0) -- (16,-32) (0,-16) -- (32,-16);
    \node[figtext, above] at (16.0,0) {Fastest, 362.7~ms\\1 of 30 pairs cross pools};
    \node[tick, below] at (8.0,-32) {Pool 1};
    \node[tick, below] at (24.0,-32) {Pool 2};
  \end{scope}
  \end{scope}
\end{tikzpicture}
    \caption{Bell-pair demand in a single QFT phase.}
    \label{fig:sharing-qft-demand}
  \end{subfigure}

  \smallskip
  \begin{subfigure}[b]{\columnwidth}
    \centering
    \begin{tikzpicture}
  \begin{axis}[
      scale only axis, width=0.86\columnwidth, height=18mm,
      xmin=0, xmax=248, ymin=-1.5, ymax=47.5,
      ytick=\empty, xlabel={Phase}, ylabel={Grouping (id)},
      axis x line*=bottom, axis y line*=left,
      tick label style={font=\fontsize{6}{7}\selectfont},
      label style={font=\fontsize{6.5}{7.5}\selectfont},
      xlabel style={yshift=3pt}, ylabel style={yshift=4pt},
      title style={font=\fontsize{6.5}{7.5}\selectfont},
      ]
    \addplot[blue, line width=0.9pt] coordinates {
      (0,0) (16,0) (16,1) (18,1) (18,2) (20,2) (20,3) (22,3)
      (22,4) (24,4) (24,5) (26,5) (26,6) (28,6) (28,7) (30,7)
      (30,8) (32,8) (32,9) (34,9) (34,10) (36,10) (36,11) (38,11)
      (38,12) (40,12) (40,13) (42,13) (42,14) (44,14) (44,15) (46,15)
      (46,16) (48,16) (48,17) (50,17) (50,18) (52,18) (52,19) (54,19)
      (54,20) (56,20) (56,21) (58,21) (58,22) (60,22) (60,23) (62,23)
      (62,24) (64,24) (64,25) (66,25) (66,26) (68,26) (68,27) (70,27)
      (70,28) (72,28) (72,29) (74,29) (74,30) (76,30) (76,7) (78,7)
      (78,31) (80,31) (80,32) (82,32) (82,33) (84,33) (84,34) (86,34)
      (86,35) (88,35) (88,36) (90,36) (90,37) (92,37) (92,38) (94,38)
      (94,39) (96,39) (96,40) (98,40) (98,41) (100,41) (100,27) (102,27)
      (102,42) (104,42) (104,43) (106,43) (106,44) (108,44) (108,27) (110,27)
      (110,45) (112,45) (112,8) (114,8) (114,46) (116,46) (116,27) (118,27)
      (118,47) (120,47) (120,0) (128,0) (128,47) (130,47) (130,27) (132,27)
      (132,46) (134,46) (134,8) (136,8) (136,45) (138,45) (138,27) (140,27)
      (140,44) (142,44) (142,43) (144,43) (144,42) (146,42) (146,27) (148,27)
      (148,41) (150,41) (150,40) (152,40) (152,39) (154,39) (154,38) (156,38)
      (156,37) (158,37) (158,36) (160,36) (160,35) (162,35) (162,34) (164,34)
      (164,33) (166,33) (166,32) (168,32) (168,31) (170,31) (170,7) (172,7)
      (172,30) (174,30) (174,29) (176,29) (176,28) (178,28) (178,27) (180,27)
      (180,26) (182,26) (182,25) (184,25) (184,24) (186,24) (186,23) (188,23)
      (188,22) (190,22) (190,21) (192,21) (192,20) (194,20) (194,19) (196,19)
      (196,18) (198,18) (198,17) (200,17) (200,16) (202,16) (202,15) (204,15)
      (204,14) (206,14) (206,13) (208,13) (208,12) (210,12) (210,11) (212,11)
      (212,10) (214,10) (214,9) (216,9) (216,8) (218,8) (218,7) (220,7)
      (220,6) (222,6) (222,5) (224,5) (224,4) (226,4) (226,3) (228,3)
      (228,2) (230,2) (230,1) (232,1) (232,0) (248,0)
    };
  \end{axis}
\end{tikzpicture}
    \caption{Fastest grouping per phase: 106 reconfigurations.}
    \label{fig:sharing-qft-sequence}
  \end{subfigure}
  \vspace{-3mm}
  \caption{Changing BSM sharing reduces conversion, but the fastest grouping changes across QFT's phases. \textit{(a)} With the same 30 Bell-pair operations and BSM capacity, regrouping reduces cross-pool pairs from 16 to one. \textit{(b)} Following the fastest grouping through 248 phases requires 106 reconfigurations, adding 10.6~s at 100~ms per grouping change.}
  \label{fig:sharing-qft}
  \vspace{-3mm}
\end{figure}

A program's communicating pairs change across phases, so the preferred grouping can change as well. We study 64-qubit QFT on 32 QPUs with two pools of four BSMs each, using the same optical parameters and path setup delay. We compare 62 balanced groupings using completion-time estimates that account for loss and contention for BSMs, links, and QPU interfaces. Choosing the fastest grouping per phase reduces cross-pool Bell-pair operations from 2,048 to 80, a 96.1\% reduction. In the single phase shown in Figure~\ref{fig:sharing-qft-demand}, regrouping reduces cross-pool pairs from 16 to one and estimated completion time from 2087.8 to 362.7~ms, with the same required operations and BSM capacity.

\myitem{Choosing the fastest grouping per phase can incur excessive reconfiguration:} A grouping's benefit must also justify the time spent establishing it. Following the fastest grouping across QFT's 248 phases requires at least 106 reconfigurations (Figure~\ref{fig:sharing-qft-sequence}). With 10~$\mu s$ of path setup per pair, estimated communication time is 39.87~s before grouping reconfiguration. At an illustrative 100~ms per grouping change (our evaluation uses 1~ms and varies it in \S\ref{sec:eval-sensitivity}), reconfiguration adds 10.6~s, increasing the total to 50.47~s. A slower grouping in some phases can be preferable when the time saved by switching does not justify the reconfiguration delay.

\smallskip
\takeaway{Reconfigurable BSM sharing can avoid conversion and balance contention without adding BSMs. Its benefits depend on both how QPUs share the pools and when that sharing changes.}

\subsection{Design Goals and Challenges}
\label{sec:design-goals}

Our goal is to reduce communication completion time by adapting how QPUs share existing BSM resources as a program executes. This requires an architecture that supports programmable sharing and a way to choose both the groupings and when they change.

\smallskip
\noindent
\textit{\textbf{(Q1)} How can the architecture enable reconfigurable BSM sharing?}

\noindent
QPUs remain in fixed racks, but the BSM pools they share need to change as a program executes. Supporting this flexibility requires programmable photonic paths that give QPUs direct access to different pools without physically relocating the processors or migrating their quantum states.

\smallskip
\noindent
\textit{\textbf{(Q2)} How should QPUs share BSM pools?}

\noindent
Programmable connections create a combinatorial space of QPU groupings: dividing just 32 QPUs evenly between two identical pools yields ${}^{32}C_{16}/2 \approx 300$ million distinct choices. Exhaustive search is impractical, so the challenge is to identify promising groupings efficiently and assess their completion times under probabilistic generation and contention for BSMs and links.

\smallskip
\noindent
\textit{\textbf{(Q3)} When should the grouping change?}

\noindent
Reconfiguration delay makes the value of a grouping depend on the phases that precede and follow it. Even after narrowing the search to $2$ groupings, the QFT example's 248 phases admit $2^{248} \approx 4.5 \times 10^{74}$ possible sequences. The challenge is to find a sequence whose communication savings justify its reconfiguration delays without exploring every sequence.

\section{The \name Architecture}
\label{sec:architecture}

The rigidity we observed in \S\ref{sec:rigid-tradeoff} comes from fixing which BSM pools QPUs can reach without conversion based on their physical rack placement. Scheduling can use BSMs elsewhere, but reaching them still requires the paths imposed by fixed connectivity. \name enables reconfigurable BSM resource sharing by allowing QPUs to change which pools they access directly. The program can use one grouping across several phases or change it as its communicating pairs change. The architecture provides these connections, while the compiler decides how and when to change them to balance photon-path loss, contention, and reconfiguration delay.

\subsection{Enabling Reconfigurable BSM Sharing}
\label{sec:arch-sharing}

As Figure~\ref{fig:architecture-red} shows, \name uses ToRs as native-wavelength optical circuit switches and places BSM pools at aggregation switches. Each ToR establishes circuits from its QPUs to selected pools. The aggregation switches connect the assigned QPUs to their BSMs, while QFC-equipped uplinks provide access to the core. QPUs in different racks can now share a pool without conversion, while QPUs in the same rack can use different pools to distribute contention. Changing these connections reshapes BSM sharing without adding BSMs, moving QPUs, or migrating quantum states.

\myitem{Reconfigurable connections across racks:} A scale-up domain in \name comprises racks connected to a common set of aggregation switches. Each ToR acts as a native-wavelength optical circuit switch, connecting its QPU-facing ports to selected aggregation uplinks. These circuits assign QPUs to BSM pools across racks and remain in place while the aggregation switches direct incoming photons to available BSMs. Reconfiguring the ToRs changes these assignments without moving QPUs or migrating quantum states, making the scale-up a reconfigurable entanglement domain. QPUs assigned to the same pool can generate Bell pairs without conversion, while pairs spanning pools use QFC-equipped uplinks and the core, even within the same scale-up domain. The ToR circuits provide the programmable access needed to realize the groupings explored in \S\ref{sec:reconfigurable-sharing}.

\myitem{Preserving native-wavelength reach:} Sharing BSMs across racks places a strict requirement on ToR insertion loss. Every photon traverses a ToR circuit, and loss at this stage both limits the scale-up native-wavelength reach and increases generation retries. Low-loss mechanical optical switches, such as MEMS-based devices, offer a way to preserve this reach, but their mechanical actuation can introduce higher reconfiguration delay. \name accepts this tradeoff to provide low-loss, programmable access to shared BSM pools.

This choice is practical because the ToR circuits need not change for every Bell pair. They connect QPUs to pools, while the aggregation switches establish paths to individual BSMs under the current grouping. A grouping can serve many operations, within a phase or across several, allowing its communication savings to offset the delay of establishing it. The question becomes how to group QPUs to benefit from this access, and when changing that grouping saves enough time to justify reconfiguration.

\subsection{Choosing QPU Groupings}
\label{sec:optimization}

The compiler must choose groupings that reduce photon-path loss without concentrating BSM contention. As we observed in~\S\ref{sec:reconfigurable-sharing}, avoiding conversion alone cannot distinguish a grouping that overloads one pool from another that allows generation to proceed concurrently. \name uses the program's communication to identify promising groupings, then compares their completion times under the available BSM and link capacity.

We represent the program as a sequence of communication phases, preserving dependencies introduced by both local and remote operations. Each remote operation requires a Bell pair between its participating QPUs. Within a phase, pairs can be generated concurrently when resources permit, and generation for the next phase begins once all required pairs are complete. For a scale-up domain with $n$ QPUs and $m$ equally provisioned pools, a grouping $g$ assigns $n/m$ QPUs to each pool without changing logical-qubit placement. The compiler constructs a candidate set $\mathcal{G}$ and estimates each phase's completion time $C_t(g)$ under every candidate, capturing both the photon paths it provides and the contention it creates.

\myitem{Finding promising groupings:} We construct a QPU interaction graph for each distinct phase, with QPUs as vertices and edge weights counting the Bell pairs required between them. Balanced graph partitioning with METIS~\cite{karypis1998multilevel,davis2023towards} groups communicating QPUs to reduce cross-pool operations while assigning the same number of QPUs to each pool. We map the resulting groups to physical pools, remove duplicate assignments, and include the original deployment grouping to form $\mathcal{G}$. A grouping that benefits one phase may also serve others efficiently, so the compiler evaluates every candidate across the full sequence. Partitioning narrows the search to groupings that can reduce conversion, while completion-time estimates account for the contention within each grouping.

\myitem{Comparing loss and contention:} For each candidate, we estimate a phase's completion time by scheduling its Bell-pair operations with the expected service times from \S\ref{sec:architecture-assumptions}. The selected photon paths determine the success probability and expected retries. Per-pair path setup adds to the time for which the BSM and links are reserved. The model also enforces the available BSM capacity, optical channels, and QPU communication interfaces, so operations sharing a resource may have to wait even when their photon paths are short. The last pair to complete determines the phase's estimated completion time. Two groupings with the same number of cross-pool pairs can have different completion times if one concentrates operations on fewer BSMs. Comparing them this way allows the compiler to weigh shorter photon paths against the waiting they may introduce.
The comparison also depends on how generation is scheduled within each grouping. As we observed in \S\ref{sec:rigid-tradeoff}, using another pool's BSM can reduce waiting, but reaching it increases loss and retries. 

\name evaluates groupings under scheduling rules that wait for local BSMs, use available remote BSMs, or allow the entanglement-swapping option used by SwitchQNet~\cite{zhang2025switchqnet}. It also considers temporarily regrouping QPUs within a phase to serve cross-pool pairs without conversion, accounting for the additional delay for each such intra-phase regrouping. This rule first serves pairs that already share a pool, then waits for active generation to complete before changing assignments for the remaining pairs. The scheduler restores the phase's selected grouping before completion, including this delay in the phase cost so that the grouping assumed at each phase boundary remains unchanged. These choices let the compiler compare reaching another pool through QFCs against reconfiguring access to avoid conversion. All rules use the same resource constraints and path-dependent generation times. We do not exploit lookahead because applying it across candidate groupings and phases makes compilation prohibitively expensive for large circuits.

\subsection{Choosing When to Reconfigure}
\label{sec:arch-sequence}

A grouping's benefit must also justify the time spent establishing it. In the QFT example from \S\ref{sec:reconfigurable-sharing}, choosing the fastest grouping in every phase leads to frequent changes whose delays accumulate over the program. \name chooses groupings across the full sequence, allowing a slower choice in some phases when switching would take more time than it saves.

For a fixed scheduling rule, let $T$ be the number of communication phases, $\mathcal{G}$ the candidate set, and $C_t(g)$ the estimated completion time of phase $t$ under grouping $g\in\mathcal{G}$. This estimate includes per-pair path setup, generation, waiting for shared resources, and any within-phase regrouping and restoration required by the scheduling rule. Let $\delta\geq 0$ be the delay incurred whenever the grouping changes between phases. The compiler selects a sequence $(g_1,\ldots,g_T)\in\mathcal{G}^T$ that minimizes
\begin{equation}
J(g_1,\ldots,g_T)
=
\sum_{t=1}^{T} C_t(g_t)
+ \delta \sum_{t=2}^{T} \mathbf{1}[g_t\neq g_{t-1}],
\label{eq:red-plan-cost}
\end{equation}
where $J$ is the estimated total communication completion time and~$\mathbf{1}[\cdot]$ equals one when its condition holds and zero otherwise. The initial grouping is configured before execution, so only subsequent changes incur reconfiguration delay.

Although there are many possible sequences, those that finish a phase in the same grouping face identical choices for the remaining phases. Keeping the least costly sequence ending in each grouping is sufficient to find the best continuation. Let $D_t(g)$ be the minimum estimated time to complete phases $1$ through $t$ and finish in grouping $g$. Since the first grouping incurs no reconfiguration delay, $D_1(g)=C_1(g)$. For $t=2,\ldots,T$, we compute
\begin{equation}
D_t(g)
=
C_t(g)+
\min_{h\in\mathcal{G}}
\left\{
D_{t-1}(h)+\delta\mathbf{1}[h\neq g]
\right\},
\label{eq:red-grouping-dp}
\end{equation}
where $h$ is the grouping used in the preceding phase. Continuing with $h=g$ incurs no reconfiguration delay, while changing from $h\neq g$ adds $\delta$. The minimum estimated completion time is $\min_{g\in\mathcal{G}}D_T(g)$. Recording the minimizing predecessor for each state and backtracking from a minimizing final grouping recovers the selected sequence.

The constant delay for every grouping change allows this recurrence to be evaluated efficiently. Define $M_{t-1}=\min_{h\in\mathcal{G}}D_{t-1}(h)$ as the minimum estimated completion time through the preceding phase. Equation~\ref{eq:red-grouping-dp} can then be written as
\begin{equation}
D_t(g)
=
C_t(g)+
\min\left\{D_{t-1}(g),\,M_{t-1}+\delta\right\}.
\label{eq:red-grouping-dp-linear}
\end{equation}
If a grouping attaining $M_{t-1}$ is already $g$, continuing without reconfiguration is no more expensive than paying $\delta$. Computing $M_{t-1}$ once per phase allows each candidate to be updated in constant time, giving $O(T|\mathcal{G}|)$ time for sequence selection after the phase costs are available. The resulting sequence minimizes the estimated objective over $\mathcal{G}$, allowing a grouping to remain in use across phases where switching offers insufficient benefit.

Expected service times make this search tractable, but probabilistic retries can change when operations complete and resources become available. Before execution, \name uses stochastic trials to compare the sequences chosen under each scheduling rule alongside promising plans that keep one grouping throughout the program without temporary regrouping. It selects the plan and scheduling rule with the lowest mean completion time, with separate trials used for evaluation. During execution, the controller follows the selected scheduling rule within each phase, including any temporary regrouping and restoration, before applying the next phase's grouping and its reconfiguration delay when needed. QPUs and stored quantum states remain in place while their access to shared BSM resources follows the selected sequence.

\subsection{Practicality}
\label{sec:practicality}

Realizing these benefits requires low-loss access to BSM pools without excessive reconfiguration delay or additional demands on QPU memory. 
The ToRs must preserve enough of the optical loss budget for native-wavelength paths to span multiple racks. Low-loss mechanical switches, including MEMS-based and beam-steering devices, can support this requirement (e.g., 0.5~dB insertion loss~\cite{keysightoptical}), but may require milliseconds to reconfigure~\cite{zhang2025switchqnet,polatis,keysightoptical}, compared to the microsecond or nanosecond switching times of quantum switches~\cite{zhao2026universal}. \name accommodates this tradeoff by keeping QPU-to-pool circuits in place wherever changing them would not pay off. The compiler accounts for insertion loss when estimating completion times and reconfiguration delay when selecting the grouping sequence. Longer reconfiguration delays favor groupings whose communication savings accumulate across more phases.

\section{Evaluation}
\label{sec:evaluation}

We evaluate whether reconfigurable BSM sharing reduces communication time, where the gains come from, and how they depend on switch loss and reconfiguration delay. We also compare \name with a hypothetical large switch and discuss the implications for memory and pair quality.

\begin{figure*}[t]
  \centering
  \begin{subfigure}[b]{0.42\textwidth}
    \centering
    \includegraphics[width=\linewidth]{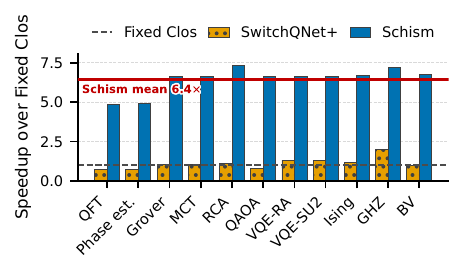}
    \caption{Speedup over Fixed Clos}
    \label{fig:eval-speedup-bars}
  \end{subfigure}\hfill
  \begin{subfigure}[b]{0.37\textwidth}
    \centering
    \includegraphics[width=\linewidth]{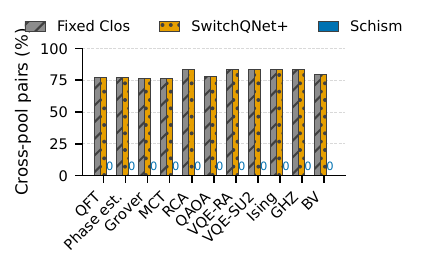}
    \caption{Cross-pool Bell pairs}
    \label{fig:eval-crosspool}
  \end{subfigure}\hfill
  \begin{subfigure}[b]{0.19\textwidth}
    \centering
    \includegraphics[width=\linewidth]{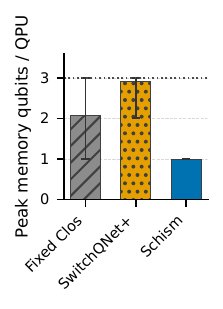}
    \caption{Memory per QPU}
    \label{fig:eval-memory}
  \end{subfigure}
  \caption{\name reduces communication time across all 11 workloads (a) by bringing communicating QPUs into the same BSM pool (b). Fixed Clos and SwitchQNet+ still generate 76--84\% of pairs across racks. These gains require less storage: \name holds at most one pair per QPU, while SwitchQNet+ fills its three-qubit memory on 10 workloads (c). Whiskers show the range across workloads.}
  \label{fig:eval-speedup}
\end{figure*}

\subsection{Methodology}
\label{sec:eval-method}

We implemented \name and all baselines in SeQUeNCe~\cite{wu2021sequence}, a discrete-event simulator for quantum networks.

\myitem{System and hardware:} Our default architecture has 32 QPUs across four racks sharing eight BSMs. Fixed Clos places two BSMs at each ToR, while \name forms four aggregation-level pools of two BSMs, each serving eight QPUs. Two core switches connect the pools. Each QPU has six data qubits, one optical interface, and three communication memory qubits for one in-flight pair and two stored pairs. The two buffer qubits follow SwitchQNet's allocation of 25\% of the eight data and buffer qubits~\cite{zhang2025switchqnet}. We model trapped-ion nodes with a $^{88}$Sr$^+$ optical interface and $^{43}$Ca$^+$ memory~\cite{stephenson2020high}. Table~\ref{tab:eval-params} lists the physical parameters. BSM switches use Bene\v{s} paths with $2\log_2 N-1$ elements for an $N\times N$ switch, giving 2.1~dB per traversal at 16 ports. \name's mechanical ToR circuits add 0.5~dB per photon and take 1~ms to regroup. Photons remain at native wavelength within a pool and undergo conversion on core paths.

\begin{table}[t]
\centering\small
\caption{Default simulation parameters for trapped-ion QPUs and the optical interconnect. ToR-circuit loss is per photon, while Bene\v{s} loss is per switching element. We vary loss and reconfiguration delay in \S\ref{sec:eval-sensitivity}.}
\label{tab:eval-params}
\begin{tabular}{@{}lr@{}}
\toprule
Parameter & Value \\
\midrule
Photon efficiency / attempt time~\cite{stephenson2020high} & 0.0209 / 1.2~$\mu s$ \\
BSM acceptance probability & 0.5 \\
Native / telecom attenuation & 30 / 0.2~dB/km \\
Loss per QFC~\cite{vanleent2022entangling} & 2.4~dB \\
Loss per Bene\v{s} element~\cite{zhao2016benes} & 0.3~dB \\
Path setup per Bell pair & 10~$\mu s$ \\
ToR-circuit loss per photon~\cite{keysightoptical} & 0.5~dB \\
Grouping reconfiguration delay & 1~ms \\
\bottomrule
\end{tabular}
\end{table}

\myitem{Comparisons:} We compare \name against \first Fixed Clos, \second SwitchQNet+, and \third BigSwitch. Fixed Clos uses BSMs at each ToR and conversion for cross-rack pairs. SwitchQNet+ runs SwitchQNet's compiler~\cite{zhang2025switchqnet} on this fabric, including lookahead, buffering, and split pairs. We extend its scheduler to wait for a BSM on a shortest path rather than take a higher-loss path through a third ToR, which performed significantly worse than Fixed Clos in our experiments. BigSwitch connects every QPU and BSM through one native-wavelength switch (\S\ref{sec:eval-bigswitch}). Two controls isolate \name's benefits: \emph{without regrouping} preserves the rack-to-pool assignments on \name's fabric, while \emph{one grouping} uses the best candidate throughout the program. The full compiler may regroup between and within phases (\S\ref{sec:optimization}).

\myitem{Workloads and metrics:} We use 11 MQT Bench circuits~\cite{quetschlich2023mqtbench}, with six data qubits per QPU: 192 qubits at 32 QPUs and 96--384 as the network grows from 16 to 64 QPUs. Qubit partitioning follows Davis et al.~\cite{davis2023towards}, with placement fixed across designs and throughout execution. Each design selects its scheduling rule using 10 simulation runs and is evaluated on 30 separate runs with independent random seeds. Speedups are ratios of mean communication completion times. We also measure cross-pool generation, memory occupancy, and regrouping overhead. Parameter experiments use workload subsets described in the captions, keeping all other parameters at their defaults.

\subsection{\name Speeds Up Every Workload}
\label{sec:eval-overall}

We first compare all 11 workloads on the same 32-QPU network. Keeping placement and BSM capacity fixed lets us relate changes in completion time to how each design serves the required Bell pairs.

\myitem{Changing BSM sharing reduces communication time:} Figure~\ref{fig:eval-speedup-bars} shows speedups of $4.9$--$7.3\times$ compared to Fixed Clos (mean $6.4\times$) and $3.6$--$8.3\times$ compared to SwitchQNet+. We call the eight workloads whose phase communication groups fit within pools of eight QPUs \emph{sparse}. QFT, phase estimation, and QAOA contain larger groups and are \emph{dense}. Sparse workloads gain $6.6$--$7.3\times$, mostly by regrouping between phases. Dense workloads also regroup within phases, achieving $6.7\times$ on QAOA and $4.9\times$ on QFT and phase estimation. Table~\ref{tab:eval-selected} gives the exact speedups and conversion counts for the workloads highlighted in the introduction.

\myitem{Regrouping removes conversion that scheduling cannot avoid:} Fixed Clos and SwitchQNet+ generate 76--84\% of pairs across racks, while \name generates every pair within a pool (Figure~\ref{fig:eval-crosspool}). This eliminates 27,705 converted pairs for QFT and 28,916 for Grover (Table~\ref{tab:eval-selected}). SwitchQNet+ can hide generation delay through lookahead, but its fixed connections preserve the need for conversion. It fills the three-qubit communication memory on 10 of 11 workloads, while \name holds at most one pair per QPU (Figure~\ref{fig:eval-memory}).

\begin{table}[t]
\centering\small
\setlength{\tabcolsep}{3pt}
\caption{\name avoids every converted pair in these workloads while spending a small fraction of communication time regrouping. Speedups compare \name with each baseline. Listed counts give converted pairs under Fixed Clos, all eliminated by \name. The upper comparison uses 32 QPUs except QAOA (64). The lower comparison separates sparse and dense workloads against BigSwitch.}
\label{tab:eval-selected}
\begin{tabular}{@{}lrrrr@{}}
\toprule
& \multicolumn{2}{c}{Speedup compared to} & Converted & Regrouping \\
Workload & Fixed Clos & SwitchQNet+ & pairs & time (\%) \\
\midrule
QFT & 4.87 & 6.73 & 27,705 & 2.54 \\
QAOA & 6.66 & 8.26 & -- & 2.03 \\
Grover & 6.61 & 6.22 & 28,916 & 0.23 \\
QAOA (64) & 7.78 & 10.63 & -- & -- \\
\midrule
\multicolumn{5}{c}{Speedup compared to BigSwitch (32 QPUs)} \\
\multicolumn{3}{l}{Eight sparse workloads} & \multicolumn{2}{r}{$1.35$--$1.38\times$} \\
\multicolumn{3}{l}{Three dense workloads} & \multicolumn{2}{r}{$0.57$--$0.89\times$} \\
\bottomrule
\end{tabular}

\end{table}

\begin{figure*}[t]
  \centering
  \begin{subfigure}[t]{0.27\textwidth}
    \centering
    \includegraphics[width=0.9\linewidth]{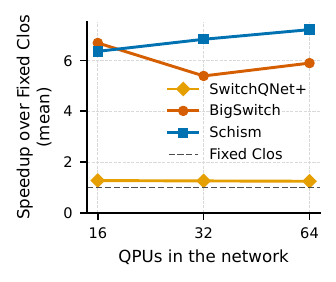}
    \caption{Network size}
    \label{fig:eval-scale}
  \end{subfigure}\hfill
  \begin{subfigure}[t]{0.42\textwidth}
    \centering
    \includegraphics[width=0.9\linewidth]{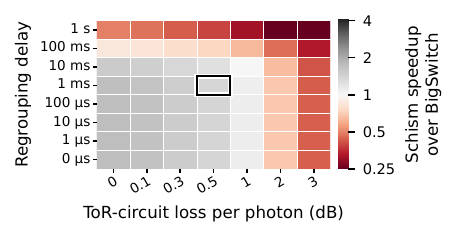}
    \caption{vs.\ BigSwitch: ToR loss $\times$ delay}
    \label{fig:eval-bigswitch-heatmap}
  \end{subfigure}\hfill
  \begin{subfigure}[t]{0.27\textwidth}
    \centering
    \includegraphics[width=0.9\linewidth]{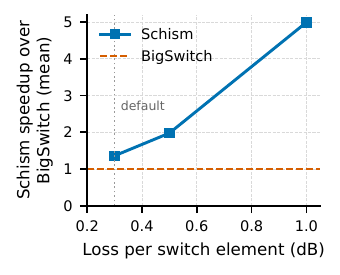}
    \caption{Fast-switch loss}
    \label{fig:eval-element-loss}
  \end{subfigure}
  \caption{Larger networks increase BigSwitch's radix and loss, allowing \name to overtake it in mean speedup (a). Higher ToR-circuit loss and regrouping delay reduce this advantage (b), while higher loss per fast switching element favors \name's smaller BSM switches (c). Panel (a) averages BV, GHZ, Ising, QAOA, RCA, VQE-RA, and VQE-SU2. Panels (b, c) average GHZ, RCA, and VQE-SU2, with the default outlined in (b).}
  \label{fig:eval-bigswitch}
  \vspace{-3mm}
\end{figure*}

\subsection{Regrouping Drives the Performance Gains}
\label{sec:eval-breakdown}

\begin{figure}[t]
  \centering
  \includegraphics[width=\linewidth]{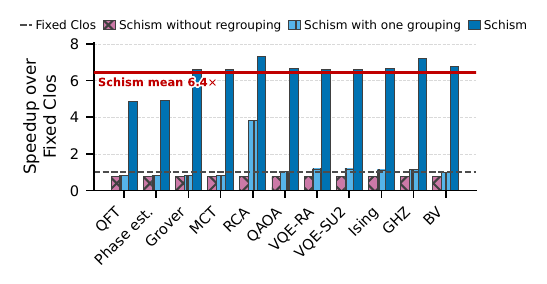}
  \vspace{-10mm}
  \caption{Adding \name's fabric without regrouping increases loss and slows communication. Choosing one grouping recovers only part of this cost, while adapting groupings during execution achieves the full gain.}
  \label{fig:eval-static}
  \vspace{-5mm}
\end{figure}

The preceding results combine the effect of \name's fabric with its grouping decisions. Figure~\ref{fig:eval-static} separates them by adding one capability at a time: the fabric without regrouping, one grouping for the whole program, and changes between and within phases. This also tests whether configuring the network once can capture the benefits of adapting to communication.

\myitem{The fabric alone adds loss:} Keeping each rack assigned to one pool preserves Fixed Clos's BSM sharing but adds a ToR circuit before the BSM switch. Its 0.5~dB loss per photon adds 1~dB across the pair, scaling joint transmission probability by $10^{-0.1}\approx0.79$. Without regrouping, \name runs at $0.79\times$ Fixed Clos's speed on every workload. Regrouping must recover this added loss before it can improve communication.

\myitem{One grouping provides limited gains:} Choosing the best candidate for the whole program improves the mean speedup to $1.26\times$, but yields only $0.80$--$1.18\times$ except on RCA ($3.9\times$), whose communication changes little. A grouping that keeps communicating QPUs together in one phase can leave them in different pools during another. Configuring connections once per program, as in designs that reconfigure slow switches between jobs~\cite{zhao2025scalable}, cannot capture the full benefit of adapting to phases.

\myitem{Changes between and within phases recover locality:} Regrouping increases the mean speedup to $6.4\times$. On dense workloads, changing only between phases leaves 15--33\% of pairs spanning pools. Within-phase changes bring those pairs into a common pool, increasing speedups from $1.7$--$2.6\times$ to $4.9$--$6.7\times$ (Table~\ref{tab:eval-within-phase}). These changes account for up to about 2.5\% of communication time, with QFT incurring the largest fraction (Table~\ref{tab:eval-selected}). The savings from avoiding conversion outweigh these delays under the default parameters.

\begin{table}[t]
\centering\small
\caption{Allowing within-phase regrouping improves on changing only between phases (PW), removing the remaining cross-pool pairs. Speedups are compared to Fixed Clos.}
\label{tab:eval-within-phase}
\begin{tabular}{@{}lrrr@{}}
\toprule
& \multicolumn{2}{c}{Speedup} & Cross-pool pairs (\%) \\
Workload & PW & \name & PW $\rightarrow$ \name \\
\midrule
QFT & 1.66 & 4.87 & $33\rightarrow 0$ \\
Phase est. & 1.70 & 4.91 & $32\rightarrow 0$ \\
QAOA & 2.60 & 6.66 & $15\rightarrow 0$ \\
\bottomrule
\end{tabular}
\end{table}

\begin{figure*}[t]
  \centering
  \begin{subfigure}[b]{0.36\textwidth}
    \centering
    \includegraphics[width=0.8\linewidth,trim=0bp 0bp 229.447bp 0bp,clip]{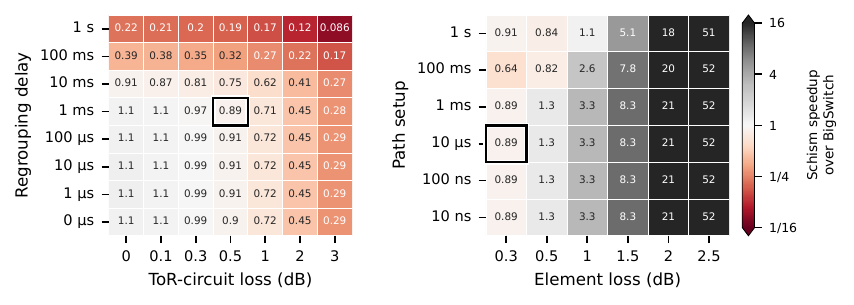}
    \caption{ToR loss (\name) and regrouping delay (QAOA)}
    \label{fig:eval-dense-tor}
  \end{subfigure}%
  \begin{subfigure}[b]{0.46\textwidth}
    \centering
    \includegraphics[width=0.8\linewidth,trim=180bp 0bp 0bp 0bp,clip]{figures/appendix/app_perpair_heatmap.pdf}
    \caption{Switch $2\times 2$ element loss and path setup (QAOA)}
    \label{fig:eval-dense-element}
  \end{subfigure}
  \begin{subfigure}[b]{0.36\textwidth}
    \centering
    \includegraphics[width=0.8\linewidth,trim=0bp 0bp 229.447bp 0bp,clip]{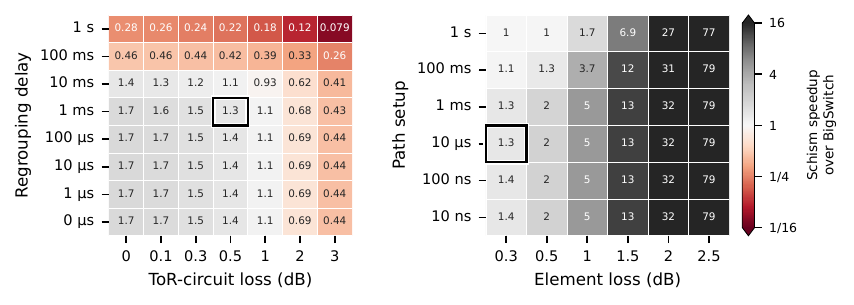}
    \caption{ToR loss (\name) and regrouping delay (GHZ)}
    \label{fig:eval-dense-tor-ghz}
  \end{subfigure}%
  \begin{subfigure}[b]{0.46\textwidth}
    \centering
    \includegraphics[width=0.8\linewidth,trim=180bp 0bp 0bp 0bp,clip]{figures/appendix/app_perpair_heatmap_ghz.pdf}
    \caption{Switch $2\times 2$ element loss and path setup (GHZ)}
    \label{fig:eval-dense-element-ghz}
  \end{subfigure}
  \caption{Comparison with BigSwitch depends on both switching tiers. As ToR-circuit loss reduces, \name outperforms BigSwitch (c), even for dense-workloads such as QAOA that requires within-phase regrouping (a). Increasing fast-switch loss favors \name's smaller BSM switches (b, d). Cells show \name's speedup compared to BigSwitch, with grey favoring \name and red favoring BigSwitch. Each panel varies its two axes while keeping the other parameters at their defaults.}
  \label{fig:eval-dense-tradeoff}
\vspace{-3mm}
\end{figure*}

\subsection{Comparison with BigSwitch}
\label{sec:eval-bigswitch}

If regrouping helps by bringing pairs into a common pool, a single pool spanning the entire network could provide this opportunity continuously. We compare with BigSwitch to understand whether avoiding grouping changes outweighs the photon loss of a larger switch.

\myitem{BigSwitch avoids conversion and regrouping:} A hypothetical network-wide Bene\v{s} switch connects every QPU to every BSM at native wavelength. We use the same 10~$\mu s$ path setup and 0.3~dB loss per element as the other BSM switches. At 32 QPUs, its $64\times64$ fabric loses 3.3~dB per photon. \name instead combines a 0.5~dB ToR circuit with a smaller BSM switch losing 2.1~dB, but pays 1~ms to regroup. On the eight sparse workloads, the lower loss outweighs regrouping and makes \name $1.35$--$1.38\times$ faster (Table~\ref{tab:eval-selected}). BigSwitch remains $1.1$--$1.75\times$ faster on the three dense workloads, where \name must also regroup within phases.

\myitem{Larger networks favor smaller BSM switches:} As network size grows from 16 to 64 QPUs, \name's mean speedup compared to Fixed Clos increases from $6.4\times$ to $7.2\times$, while BigSwitch's falls from $6.7\times$ to $5.4$--$5.9\times$ as its radix and loss grow (Figure~\ref{fig:eval-scale}). \name overtakes BigSwitch in the mean by 32 QPUs. At 64 QPUs, QAOA gains $7.78\times$ compared to Fixed Clos and $10.63\times$ compared to SwitchQNet+ (Table~\ref{tab:eval-selected}). Higher loss per switching element also favors \name: increasing it from 0.3 to 1~dB increases its advantage over BigSwitch from $1.36\times$ to $5.0\times$ on GHZ, RCA, and VQE-SU2 (Figure~\ref{fig:eval-element-loss}). Conversely, higher ToR-circuit loss or slower regrouping favors BigSwitch. On these workloads, \name leads with up to 1~dB ToR loss and 1~ms regrouping, or 10~ms at 0.5~dB, while BigSwitch wins from 100~ms (Figure~\ref{fig:eval-bigswitch-heatmap}).

\myitem{Dense workloads expose the cost of ToR circuits:} QAOA benefits from avoiding conversion but still trails BigSwitch at the default parameters. Figures~\ref{fig:eval-dense-tor},~\ref{fig:eval-dense-element} separate the effects of regrouping delay and switch loss. Even with instantaneous regrouping, \name reaches only $0.90\times$ BigSwitch's speed at 0.5~dB ToR loss. At the default 1~ms delay, reducing this loss to 0.3~dB brings it close ($0.97\times$), and 0.1~dB or less allows it to lead. Increasing loss per fast switching element also favors \name, whose smaller switches traverse fewer elements. At 10~$\mu s$ path setup, \name overtakes BigSwitch from 0.5~dB per element. The dense-workload gap depends on both technologies, rather than regrouping delay alone. For sparse workloads, such as GHZ (Figures~\ref{fig:eval-dense-tor-ghz},~\ref{fig:eval-dense-element-ghz}), \name outperforms BigSwitch even at larger reconfiguration delays, and higher ToR-loss.

\subsection{Sensitivity to Loss and Delay}
\label{sec:eval-sensitivity}

We next vary the costs that determine whether regrouping pays off: ToR-circuit loss, reconfiguration delay, and conversion loss.

\begin{figure}[t]
  \centering
  \begin{subfigure}[t]{0.55\linewidth}
    \centering
    \includegraphics[width=\linewidth]{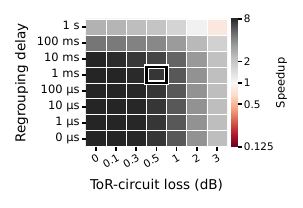}
    \vspace{-7mm}
    \caption{vs.\ Fixed Clos: ToR loss $\times$ delay}
    \label{fig:eval-access}
  \end{subfigure}\hfill
  \begin{subfigure}[t]{0.45\linewidth}
    \centering
    \includegraphics[width=\linewidth]{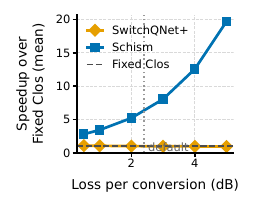}
    \vspace{-7mm}
    \caption{Conversion loss}
    \label{fig:eval-qfc}
  \end{subfigure}
  \vspace{-3mm}
  \caption{\name remains faster than Fixed Clos as ToR-circuit loss rises to 3~dB and regrouping delay to 100~ms, averaged over GHZ, RCA, and VQE-SU2 (a). Grey cells favor \name, red cells favor Fixed Clos, and the outline marks the default. Higher conversion loss increases the benefit of avoiding cross-pool paths, as shown by the mean over QFT, QAOA, VQE-SU2, and Ising (b).}
  \label{fig:eval-sensitivity}
  \vspace{-5mm}
\end{figure}

\myitem{The gains tolerate additional ToR-circuit loss:} Every photon in \name traverses a ToR circuit, whose loss increases retries. Across GHZ, RCA, and VQE-SU2, mean speedup compared to Fixed Clos falls from $9.0\times$ with lossless circuits to $2.3\times$ at 3~dB per photon, but remains above one through 100~ms regrouping delay (Figure~\ref{fig:eval-access}). At 1~s, the mean remains $1.1$--$2.6\times$ for losses up to 2~dB and falls below one at 3~dB. The benefits depend on recovering both the added photon loss and the time spent changing groupings.

\myitem{Choosing when to regroup matters as delay grows:} Under the trapped-ion parameters, local generation takes tens of milliseconds, so millisecond regrouping can be recovered across subsequent operations. At 0.5~dB ToR loss, the mean speedup remains $3.9\times$ even with 100~ms regrouping. Table~\ref{tab:eval-sequence} isolates the value of sequence selection by comparing \name with independently choosing each phase's fastest grouping. This alternative uses the same candidates and estimates, but ignores between-phase regrouping delay during selection and does not consider keeping one grouping throughout. It is within 2\% of \name at 1~ms, but takes $1.2$--$1.6\times$ longer at 100~ms on GHZ, RCA, and VQE-SU2. At 1~s, \name keeps one grouping on these workloads, while the alternative spends 80--98\% of its time regrouping. On QAOA, both approaches choose the same sequence through 100~ms, showing that the value of delay-aware selection also depends on communication structure.

\begin{table}[t]
\centering\small
\caption{Accounting for delay avoids costly changes as regrouping slows. Ratios divide the fastest-per-phase choice's time by \name's at 0.5~dB ToR loss. The last columns show \name's changes and the per-phase choice's regrouping-time share at 1~s.}
\label{tab:eval-sequence}
\setlength{\tabcolsep}{3pt}
\begin{tabular}{@{}lrrrrr@{}}
\toprule
& \multicolumn{3}{c}{Completion-time ratio} & \multicolumn{2}{c}{At 1~s} \\
Workload & 1~ms & 100~ms & 1~s & Changes & Time (\%) \\
\midrule
GHZ & 1.02 & 1.60 & 6.85 & 0 & 98 \\
RCA & 1.00 & 1.21 & 2.71 & 0 & 80 \\
VQE-SU2 & 1.01 & 1.35 & 3.43 & 0 & 95 \\
QAOA & 1.00 & 1.00 & 1.05 & 9 & 47 \\
\bottomrule
\end{tabular}
\end{table}

\myitem{Conversion loss amplifies the benefit:} Fixed Clos requires two conversions per cross-rack pair, while \name generates every pair within a pool in these experiments. Increasing loss per QFC from 0.5 to 5~dB increases \name's mean speedup from $2.8\times$ to $20\times$ (Figure~\ref{fig:eval-qfc}). SwitchQNet+ remains close to Fixed Clos in the mean because scheduling cannot remove the conversion imposed by fixed connectivity. Even at the lowest loss tested, avoiding core paths and reducing contention preserve substantial gains.

\subsection{Discussion and Limitations}
\label{sec:eval-quality}
We discuss the implications of reconfigurable BSM sharing for memory and pair quality, and the limits of our evaluation.

\myitem{Scope of our results:} We measure the time to generate heralded Bell pairs under the selected trapped-ion parameters, without evaluating fidelity. This isolates the effects of BSM sharing, photon-path loss, and contention. The experiments show where these gains persist and where switch loss or delay outweighs them. Pair quality requires platform-specific memory and gate models beyond the completion-time objective.

\myitem{Regrouping requires no reserved buffer:} \name holds at most one pair per QPU in these experiments and reserves no cache for future operations. Generated pairs still occupy qubits until consumed, and pairs completed earlier in a phase may wait for the remaining pairs. Avoiding speculative generation reduces storage requirements, but fidelity depends on these waiting times. Lookahead could complement regrouping when memory permits.

\myitem{Faster communication shortens the wait for quantum states:} Data qubits must remain coherent throughout generation and reconfiguration. \name reduces that communication interval, while its compiler already accounts for time spent regrouping. Swapping and purification add local operations, storage, and elementary pairs~\cite{dawar2026,azuma2023quantum}. Faster generation can reduce the time needed to obtain these pairs, but local-operation errors and memory decay also affect their quality. Incorporating these costs would let the compiler compare groupings by the time needed to produce pairs at the fidelity required by the program.

\section{Related Work}
\label{sec:related_work}

\name connects reconfigurable interconnects with distributed quantum compilation. Prior work shapes the communication a program needs or schedules it over available resources. Our focus is on adapting which QPUs share the BSM pools that serve this communication.

\myitem{Quantum datacenter architectures:} Photonic links connect modular quantum registers~\cite{monroe2014large}, while dynamic quantum networks extend resource sharing toward datacenter and cloud infrastructure~\cite{nejabati2022dynamic}. Switch-centric and server-centric designs study how topology and physical-layer constraints affect distributed execution~\cite{shapourian2025quantum}. Quantum-information-preserving switches support routing without destroying the photons' encoded states~\cite{zhao2026universal}. \name builds on this setting by making native-wavelength connections to shared BSM pools reconfigurable during execution.

\myitem{Entanglement routing and scheduling:} Routing techniques select paths and swapping operations under probabilistic generation and limited memory~\cite{sanguard,dawar2026}. They include multipath distribution~\cite{pant2019routing}, pre-generated and on-demand entanglement~\cite{chakraborty2020distributed}, and concurrent source-destination pairs~\cite{shi2020concurrent}. Program-derived demand guides resource allocation and switch control~\cite{vista2025entanglement}, while network-aware scheduling accounts for generation probabilities, communication-qubit availability, switch delays, and memory coherence~\cite{pouryousef2025networkaware}. SwitchQNet coordinates program and network scheduling~\cite{zhang2025switchqnet}. \name changes the QPU-to-BSM assignments under which these scheduling decisions operate, allowing the program to avoid converted paths as its communication changes.

\myitem{Circuit partitioning and qubit placement:} Graph and hypergraph partitioning reduce communication across QPUs~\cite{PhysRevA.100.032308,davis2023towards}, including heterogeneous and topology-aware mappings~\cite{Andres-Martinez_2024}. Time-sliced placement accounts for moving quantum data between mappings~\cite{baker2020time}, while joint allocation and scheduling consider network resources and decoherence~\cite{sundaram2024distributed}. These techniques determine which QPUs must communicate. \name applies partitioning after logical-qubit placement: its graph vertices are QPUs, and its groups determine shared access to BSM pools. It compares these groupings using completion-time estimates that account for loss and contention, without migrating quantum states.

\myitem{Reconfigurable classical datacenters:} Optical circuit switching has been active research in the recent past~\cite{NANCEHALL2021100621,10.1145/3708980,10.1145/1851182.1851223,10.1145/2486001.2486007,photonicsreconfiguration,10.1145/2716281.2836126,10.1145/2619239.2626328,10.1145/2896377.2901479,10.1145/2934872.2934911,10.1145/3387514.3406221,10.1145/3651890.3672248,10.1145/3579312,10.1145/3098822.3098838,vermilion,10.1145/3718958.3750465,10.1145/3696348.3696856}, spanning both demand-oblivious configurations~\cite{10.1145/3098822.3098838,10.1145/3387514.3406221,10.1145/3579312,10.1145/3651890.3672248,pompili2021realization} and demand-aware schedules that optimize throughput or completion time~\cite{10.1145/1851182.1851223,10.1145/2896377.2901479,10.1145/2716281.2836126,vermilion,amponsah2026tetris}. RDC~\cite{278374} proposes reconfigurability at the edge using an additional optical switch below the electrical ToR switches in the context of packet switching and bandwidth utilization across ToRs. Recent designs adapt connectivity to AI communication~\cite{10.1145/3748273.3749203,rahman2026harvest,10.1145/3772356.3772395,10.1145/3779212.3790238,10.1145/3789240.3829187,10.1145/3712285.3759842}. \name also balances communication savings against reconfiguration delay, but its configurations allocate shared BSM resources. Importantly, \name's decisions depend on photon loss, probabilistic generation, and contention, beyond traffic volume and link bandwidth.
\section{Conclusion}

We presented \name, an architecture and compiler that enable reconfigurable BSM sharing across racks without moving QPUs or migrating quantum states. On 32 QPUs with trapped-ion parameters, \name reduces communication time by $4.9$--$7.3\times$ (mean $6.4\times$) over Fixed Clos and by $3.6$--$8.3\times$ over SwitchQNet+ on all 11 workloads, while regrouping for up to about 2.5\% of the run. We showed that changing which QPUs share each pool can reduce wavelength conversion and BSM contention, while choosing when to change these assignments balances communication savings against reconfiguration delay. These benefits come from adapting access to existing BSM resources, without increasing their capacity. This work opens opportunities to jointly optimize qubit placement and BSM sharing, and to coordinate access across concurrently executing programs, extending reconfigurable resource sharing to the broader demands of a quantum datacenter.

\begin{acks}
We used GPT-6 Astra and Claude Opus 5.5 to improve the grammar and readability of text initially written by the authors and to assist with coding. All ideas and intellectual contributions originated with the authors. We reviewed and verified all final text.
\end{acks}

\bibliographystyle{ACM-Reference-Format}
\bibliography{references}

@article{sanguard,
  title = {Network requirements for distributed quantum computation},
  author = {Jacinto, Hugo and Gouzien, {\'E}lie and Sangouard, Nicolas},
  journal = {Phys. Rev. Res.},
  volume = {8},
  issue = {1},
  pages = {013205},
  numpages = {7},
  year = {2026},
  month = {Feb},
  publisher = {American Physical Society},
  doi = {10.1103/v9ln-c4v2},
  url = {https://link.aps.org/doi/10.1103/v9ln-c4v2}
}

@inproceedings{10.1145/3651890.3672248,
author = {Amir, Daniel and Saran, Nitika and Wilson, Tegan and Kleinberg, Robert and Shrivastav, Vishal and Weatherspoon, Hakim},
title = {Shale: A Practical, Scalable Oblivious Reconfigurable Network},
year = {2024},
isbn = {9798400706141},
publisher = {Association for Computing Machinery},
address = {New York, NY, USA},
url = {https://doi.org/10.1145/3651890.3672248},
doi = {10.1145/3651890.3672248},
booktitle = {Proceedings of the ACM SIGCOMM 2024 Conference},
pages = {449–464},
numpages = {16},
location = {Sydney, NSW, Australia},
series = {ACM SIGCOMM '24}
}

@inproceedings{10.1145/2486001.2486007,
author = {Porter, George and Strong, Richard and Farrington, Nathan and Forencich, Alex and Chen-Sun, Pang and Rosing, Tajana and Fainman, Yeshaiahu and Papen, George and Vahdat, Amin},
title = {Integrating microsecond circuit switching into the data center},
year = {2013},
isbn = {9781450320566},
publisher = {Association for Computing Machinery},
address = {New York, NY, USA},
url = {https://doi.org/10.1145/2486001.2486007},
doi = {10.1145/2486001.2486007},
booktitle = {Proceedings of the ACM SIGCOMM 2013 Conference on SIGCOMM},
pages = {447–458},
numpages = {12},
location = {Hong Kong, China},
series = {SIGCOMM '13}
}

@inproceedings{10.1145/1851182.1851223,
author = {Farrington, Nathan and Porter, George and Radhakrishnan, Sivasankar and Bazzaz, Hamid Hajabdolali and Subramanya, Vikram and Fainman, Yeshaiahu and Papen, George and Vahdat, Amin},
title = {Helios: a hybrid electrical/optical switch architecture for modular data centers},
year = {2010},
isbn = {9781450302012},
publisher = {Association for Computing Machinery},
address = {New York, NY, USA},
url = {https://doi.org/10.1145/1851182.1851223},
doi = {10.1145/1851182.1851223},
booktitle = {Proceedings of the ACM SIGCOMM 2010 Conference},
pages = {339–350},
numpages = {12},
location = {New Delhi, India},
series = {SIGCOMM '10}
}

@article{10.1145/3708980,
author = {Avin, Chen and Schmid, Stefan},
title = {Revolutionizing Datacenter Networks via Reconfigurable Topologies},
year = {2025},
issue_date = {June 2025},
publisher = {Association for Computing Machinery},
address = {New York, NY, USA},
volume = {68},
number = {6},
issn = {0001-0782},
url = {https://doi.org/10.1145/3708980},
doi = {10.1145/3708980},
journal = {Commun. ACM},
month = jun,
pages = {44–53},
numpages = {10}
}

@inproceedings{vermilion,
  author = {Addanki, Vamsi and Rao, Deeksha P and Knabe, Goran Dario and Avin, Chen and Schmid, Stefan},
  title = {Vermilion: A Traffic-Aware Reconfigurable Optical Interconnect with Formal Throughput Guarantees},
  booktitle = {24th USENIX Symposium on Networked Systems Design and Implementation (NSDI 27)},
  year = {2027},
  address = {Providence, RI},
  publisher = {USENIX Association},
  url = {https://www.usenix.org/conference/nsdi27/presentation/addanki}
}

@article{NANCEHALL2021100621,
title = {A Survey of Reconfigurable Optical Networks},
journal = {Optical Switching and Networking},
volume = {41},
pages = {100621},
year = {2021},
issn = {1573-4277},
doi = {https://doi.org/10.1016/j.osn.2021.100621},
url = {https://www.sciencedirect.com/science/article/pii/S1573427721000187},
author = {Matthew {Nance Hall} and Klaus-Tycho Foerster and Stefan Schmid and Ramakrishnan Durairajan}
}

@inproceedings{10.1145/3779212.3790238,
author = {Vijaya Kumar, Abhishek and Ding, Eric and Devraj, Arjun and Bunandar, Darius and Singh, Rachee},
title = {Reconfigurable Torus Fabrics for Multi-tenant ML},
year = {2026},
isbn = {9798400723599},
publisher = {Association for Computing Machinery},
address = {New York, NY, USA},
url = {https://doi.org/10.1145/3779212.3790238},
doi = {10.1145/3779212.3790238},
booktitle = {Proceedings of the 31st ACM International Conference on Architectural Support for Programming Languages and Operating Systems, Volume 2},
pages = {1547–1565},
numpages = {19},
location = {USA},
series = {ASPLOS '26}
}

@inproceedings{10.1145/3696348.3696856,
author = {Kumar, Abhishek Vijaya and Devraj, Arjun and Bunandar, Darius and Singh, Rachee},
title = {A case for server-scale photonic connectivity},
year = {2024},
isbn = {9798400712722},
publisher = {Association for Computing Machinery},
address = {New York, NY, USA},
url = {https://doi.org/10.1145/3696348.3696856},
doi = {10.1145/3696348.3696856},
booktitle = {Proceedings of the 23rd ACM Workshop on Hot Topics in Networks},
pages = {290–299},
numpages = {10},
location = {Irvine, CA, USA},
series = {HotNets '24}
}

@inproceedings{10.1145/3098822.3098838,
author = {Mellette, William M. and McGuinness, Rob and Roy, Arjun and Forencich, Alex and Papen, George and Snoeren, Alex C. and Porter, George},
title = {RotorNet: A Scalable, Low-complexity, Optical Datacenter Network},
year = {2017},
isbn = {9781450346535},
publisher = {Association for Computing Machinery},
address = {New York, NY, USA},
url = {https://doi.org/10.1145/3098822.3098838},
doi = {10.1145/3098822.3098838},
booktitle = {Proceedings of the Conference of the ACM Special Interest Group on Data Communication},
pages = {267–280},
numpages = {14},
location = {Los Angeles, CA, USA},
series = {SIGCOMM '17}
}

@inproceedings{10.1145/3387514.3406221,
author = {Ballani, Hitesh and Costa, Paolo and Behrendt, Raphael and Cletheroe, Daniel and Haller, Istvan and Jozwik, Krzysztof and Karinou, Fotini and Lange, Sophie and Shi, Kai and Thomsen, Benn and Williams, Hugh},
title = {Sirius: A Flat Datacenter Network with Nanosecond Optical Switching},
year = {2020},
isbn = {9781450379557},
publisher = {Association for Computing Machinery},
address = {New York, NY, USA},
url = {https://doi.org/10.1145/3387514.3406221},
doi = {10.1145/3387514.3406221},
booktitle = {Proceedings of the Annual Conference of the ACM Special Interest Group on Data Communication on the Applications, Technologies, Architectures, and Protocols for Computer Communication},
pages = {782–797},
numpages = {16},
location = {Virtual Event, USA},
series = {SIGCOMM '20}
}

@article{10.1145/3579312,
author = {Addanki, Vamsi and Avin, Chen and Schmid, Stefan},
title = {Mars: Near-Optimal Throughput with Shallow Buffers in Reconfigurable Datacenter Networks},
year = {2023},
issue_date = {March 2023},
publisher = {Association for Computing Machinery},
address = {New York, NY, USA},
volume = {7},
number = {1},
url = {https://doi.org/10.1145/3579312},
doi = {10.1145/3579312},
journal = {Proc. ACM Meas. Anal. Comput. Syst.},
month = {mar},
articleno = {2},
numpages = {43}
}

@inproceedings{10.1145/2934872.2934911,
author = {Ghobadi, Monia and Mahajan, Ratul and Phanishayee, Amar and Devanur, Nikhil and Kulkarni, Janardhan and Ranade, Gireeja and Blanche, Pierre-Alexandre and Rastegarfar, Houman and Glick, Madeleine and Kilper, Daniel},
title = {ProjecToR: Agile Reconfigurable Data Center Interconnect},
year = {2016},
isbn = {9781450341936},
publisher = {Association for Computing Machinery},
address = {New York, NY, USA},
url = {https://doi.org/10.1145/2934872.2934911},
doi = {10.1145/2934872.2934911},
booktitle = {Proceedings of the 2016 ACM SIGCOMM Conference},
pages = {216–229},
numpages = {14},
location = {Florianopolis, Brazil},
series = {SIGCOMM '16}
}

@inproceedings{10.1145/1402958.1402967,
author = {Al-Fares, Mohammad and Loukissas, Alexander and Vahdat, Amin},
title = {A scalable, commodity data center network architecture},
year = {2008},
isbn = {9781605581750},
publisher = {Association for Computing Machinery},
address = {New York, NY, USA},
url = {https://doi.org/10.1145/1402958.1402967},
doi = {10.1145/1402958.1402967},
booktitle = {Proceedings of the ACM SIGCOMM 2008 Conference on Data Communication},
pages = {63–74},
numpages = {12},
location = {Seattle, WA, USA},
series = {SIGCOMM '08}
}

@inproceedings{10.1145/2619239.2626328,
author = {Hamedazimi, Navid and Qazi, Zafar and Gupta, Himanshu and Sekar, Vyas and Das, Samir R. and Longtin, Jon P. and Shah, Himanshu and Tanwer, Ashish},
title = {FireFly: a reconfigurable wireless data center fabric using free-space optics},
year = {2014},
isbn = {9781450328364},
publisher = {Association for Computing Machinery},
address = {New York, NY, USA},
url = {https://doi.org/10.1145/2619239.2626328},
doi = {10.1145/2619239.2626328},
booktitle = {Proceedings of the 2014 ACM Conference on SIGCOMM},
pages = {319–330},
numpages = {12},
location = {Chicago, Illinois, USA},
series = {SIGCOMM '14}
}

@inproceedings{10.1145/2896377.2901479,
author = {Bojja Venkatakrishnan, Shaileshh and Alizadeh, Mohammad and Viswanath, Pramod},
title = {Costly Circuits, Submodular Schedules and Approximate Carath\'{e}odory Theorems},
year = {2016},
isbn = {9781450342667},
publisher = {Association for Computing Machinery},
address = {New York, NY, USA},
url = {https://doi.org/10.1145/2896377.2901479},
doi = {10.1145/2896377.2901479},
booktitle = {Proceedings of the 2016 ACM SIGMETRICS International Conference on Measurement and Modeling of Computer Science},
pages = {75–88},
numpages = {14},
location = {Antibes Juan-les-Pins, France},
series = {SIGMETRICS '16}
}

@inproceedings{10.1145/2716281.2836126,
author = {Liu, He and Mukerjee, Matthew K. and Li, Conglong and Feltman, Nicolas and Papen, George and Savage, Stefan and Seshan, Srinivasan and Voelker, Geoffrey M. and Andersen, David G. and Kaminsky, Michael and Porter, George and Snoeren, Alex C.},
title = {Scheduling techniques for hybrid circuit/packet networks},
year = {2015},
isbn = {9781450334129},
publisher = {Association for Computing Machinery},
address = {New York, NY, USA},
url = {https://doi.org/10.1145/2716281.2836126},
doi = {10.1145/2716281.2836126},
booktitle = {Proceedings of the 11th ACM Conference on Emerging Networking Experiments and Technologies},
articleno = {41},
numpages = {13},
location = {Heidelberg, Germany},
series = {CoNEXT '15}
}

@misc{polatis,
  author={{Polatis}},
  title={{POLATIS® 7000 Optical Circuit Switch}},
  note={\url{https://www.hubersuhner.com/en/shop/product/other-systems/optical-switches/rack-mount-circuit-switches/85223159/polatis-7000-optical-circuit-switch}}
}

@inproceedings{10.1145/3748273.3749203,
author = {Athapathu, Rukshani and Porter, George},
title = {Reconfigurability within Collective Communication Algorithms},
year = {2025},
isbn = {9798400720826},
publisher = {Association for Computing Machinery},
address = {New York, NY, USA},
url = {https://doi.org/10.1145/3748273.3749203},
doi = {10.1145/3748273.3749203},
booktitle = {Proceedings of the 2nd Workshop on Networks for AI Computing},
pages = {43–49},
numpages = {7},
location = {Coimbra, Portugal},
series = {NAIC '25}
}

@inproceedings{10.1145/3712285.3759842,
author = {Wu, Zhenguo and Klenk, Benjamin and Dennison, Larry and Bergman, Keren},
title = {ACTINA: Adapting Circuit-Switching Techniques for AI Networking Architectures},
year = {2025},
isbn = {9798400714665},
publisher = {Association for Computing Machinery},
address = {New York, NY, USA},
url = {https://doi.org/10.1145/3712285.3759842},
doi = {10.1145/3712285.3759842},
booktitle = {Proceedings of the International Conference for High Performance Computing, Networking, Storage and Analysis},
pages = {1211–1222},
numpages = {12},
location = {
},
series = {SC '25}
}

@inproceedings{10.1145/3772356.3772395,
author = {Addanki, Vamsi},
title = {When Light Bends to the Collective Will: A Theory and Vision for Adaptive Photonic Scale-up Domains},
year = {2025},
isbn = {9798400722806},
publisher = {Association for Computing Machinery},
address = {New York, NY, USA},
url = {https://doi.org/10.1145/3772356.3772395},
doi = {10.1145/3772356.3772395},
booktitle = {Proceedings of the 24th ACM Workshop on Hot Topics in Networks},
pages = {326–334},
numpages = {9},
location = {UMD Campus, College Park, MD, USA},
series = {HotNets '25}
}

@inproceedings{10.1145/3718958.3750465,
author = {Liao, Xudong and Sun, Yijun and Tian, Han and Wan, Xinchen and Jin, Yilun and Wang, Zilong and Ren, Zhenghang and Huang, Xinyang and Li, Wenxue and Tse, Kin Fai and Zhong, Zhizhen and Liu, Guyue and Zhang, Ying and Ye, Xiaofeng and Zhang, Yiming and Chen, Kai},
title = {MixNet: A Runtime Reconfigurable Optical-Electrical Fabric for Distributed Mixture-of-Experts Training},
year = {2025},
isbn = {9798400715242},
publisher = {Association for Computing Machinery},
address = {New York, NY, USA},
url = {https://doi.org/10.1145/3718958.3750465},
doi = {10.1145/3718958.3750465},
booktitle = {Proceedings of the ACM SIGCOMM 2025 Conference},
pages = {554–574},
numpages = {21},
location = {S\~{a}o Francisco Convent, Coimbra, Portugal},
series = {SIGCOMM '25}
}

@inbook{10.1145/3789240.3829187,
author = {Ding, Eric and Lyu, Barry and Kataria, Bhaskar and Singh, Rachee},
title = {Opus: Photonic Rail-Optimized Fabric in ML Datacenters},
year = {2026},
isbn = {9798400724671},
publisher = {Association for Computing Machinery},
address = {New York, NY, USA},
url = {https://doi.org/10.1145/3789240.3829187},
booktitle = {Proceedings of the ACM SIGCOMM 2026 Conference},
pages = {922–939},
numpages = {18}
}

@inproceedings{zhang2025switchqnet,
author = {Zhang, Hezi and Xu, Yiran and Hu, Haotian and Yin, Keyi and Shapourian, Hassan and Zhao, Jiapeng and Kompella, Ramana Rao and Nejabati, Reza and Ding, Yufei},
title = {SwitchQNet: Optimizing Distributed Quantum Computing for Quantum Data Centers with Switch Networks},
year = {2025},
isbn = {9798400712616},
publisher = {Association for Computing Machinery},
address = {New York, NY, USA},
url = {https://doi.org/10.1145/3695053.3731046},
doi = {10.1145/3695053.3731046},
booktitle = {Proceedings of the 52nd Annual International Symposium on Computer Architecture},
pages = {1449–1463},
numpages = {15},
location = {Tokyo, Japan},
series = {ISCA '25}
}

@misc{shapourian2025quantum,
      title={Quantum Data Center Infrastructures: A Scalable Architectural Design Perspective}, 
      author={Hassan Shapourian and Eneet Kaur and Troy Sewell and Jiapeng Zhao and Michael Kilzer and Ramana Kompella and Reza Nejabati},
      year={2025},
      eprint={2501.05598},
      archivePrefix={arXiv},
      primaryClass={quant-ph},
      url={https://arxiv.org/abs/2501.05598}, 
}

@inproceedings {278374,
author = {Weitao Wang and Dingming Wu and Sushovan Das and Afsaneh Rahbar and Ang Chen and T. S. Eugene Ng},
title = {{RDC}: {Energy-Efficient} Data Center Network Congestion Relief with Topological Reconfigurability at the Edge},
booktitle = {19th USENIX Symposium on Networked Systems Design and Implementation (NSDI 22)},
year = {2022},
isbn = {978-1-939133-27-4},
address = {Renton, WA},
pages = {1267--1288},
url = {https://www.usenix.org/conference/nsdi22/presentation/wang-weitao-rdc},
publisher = {USENIX Association},
month = apr
}

@article{karypis1998multilevel,
author = {Karypis, George and Kumar, Vipin},
title = {A Fast and High Quality Multilevel Scheme for Partitioning Irregular Graphs},
journal = {SIAM Journal on Scientific Computing},
volume = {20},
number = {1},
pages = {359-392},
year = {1998},
doi = {10.1137/S1064827595287997},

URL = { 
    
        https://doi.org/10.1137/S1064827595287997
    
    

},
eprint = { 
    
        https://doi.org/10.1137/S1064827595287997
    
    

}
}

@PREAMBLE{
 "\providecommand{\noopsort}[1]{}" 
 # "\providecommand{\singleletter}[1]{#1}%" 
}

@article{pant2019routing,
  title={Routing entanglement in the quantum internet},
  author={Pant, Mihir and Krovi, Hari and Towsley, Don and Tassiulas, Leandros and Jiang, Liang and Basu, Prithwish and Englund, Dirk and Guha, Saikat},
  journal={npj Quantum Information},
  volume={5},
  number={1},
  pages={25},
  year={2019},
  publisher={Nature Publishing Group UK London},
  doi={10.1038/s41534-019-0139-x},
  url={https://www.nature.com/articles/s41534-019-0139-x}
}

@article{dawar2026,
 title = {Resource estimation for entanglement routing in quantum networks},
 author = {Dawar, Manik and Riedinger, Ralf and Vyas, Nilesh and Mendes, Paulo},
 journal = {Phys. Rev. A},
 volume = {113},
 issue = {2},
 pages = {022622},
 numpages = {14},
 year = {2026},
 month = {Feb},
 publisher = {American Physical Society},
 doi = {10.1103/vf4x-rjsd},
 url = {https://link.aps.org/doi/10.1103/vf4x-rjsd}
}

@article{azuma2023quantum,
  title={Quantum repeaters: From quantum networks to the quantum internet},
  author={Azuma, Koji and Economou, Sophia E and Elkouss, David and Hilaire, Paul and Jiang, Liang and Lo, Hoi-Kwong and Tzitrin, Ilan},
  journal={Reviews of Modern Physics},
  volume={95},
  number={4},
  pages={045006},
  year={2023},
  publisher={APS}
}

@article{stephenson2020high,
  title={High-rate, high-fidelity entanglement of qubits across an elementary quantum network},
  author={Stephenson, LJ and Nadlinger, DP and Nichol, BC and An, S and Drmota, P and Ballance, TG and Thirumalai, K and Goodwin, JF and Lucas, DM and Ballance, CJ},
  journal={Physical review letters},
  volume={124},
  number={11},
  pages={110501},
  year={2020},
  publisher={APS}
}

@article{pompili2021realization,
  title={Realization of a multinode quantum network of remote solid-state qubits},
  author={Pompili, Matteo and Hermans, Sophie LN and Baier, Simon and Beukers, Hans KC and Humphreys, Peter C and Schouten, Raymond N and Vermeulen, Raymond FL and Tiggelman, Marijn J and dos Santos Martins, Laura and Dirkse, Bas and others},
  journal={Science},
  volume={372},
  number={6539},
  pages={259--264},
  year={2021},
  publisher={American Association for the Advancement of Science}
}

@inproceedings{baker2020time,
author = {Baker, Jonathan M. and Duckering, Casey and Hoover, Alexander and Chong, Frederic T.},
title = {Time-sliced quantum circuit partitioning for modular architectures},
year = {2020},
isbn = {9781450379564},
publisher = {Association for Computing Machinery},
address = {New York, NY, USA},
url = {https://doi.org/10.1145/3387902.3392617},
doi = {10.1145/3387902.3392617},
booktitle = {Proceedings of the 17th ACM International Conference on Computing Frontiers},
pages = {98–107},
numpages = {10},
location = {Catania, Sicily, Italy},
series = {CF '20}
}

@inproceedings{shi2020concurrent,
author = {Shi, Shouqian and Qian, Chen},
title = {Concurrent Entanglement Routing for Quantum Networks: Model and Designs},
year = {2020},
isbn = {9781450379557},
publisher = {Association for Computing Machinery},
address = {New York, NY, USA},
url = {https://doi.org/10.1145/3387514.3405853},
doi = {10.1145/3387514.3405853},
booktitle = {Proceedings of the Annual Conference of the ACM Special Interest Group on Data Communication on the Applications, Technologies, Architectures, and Protocols for Computer Communication},
pages = {62–75},
numpages = {14},
location = {Virtual Event, USA},
series = {SIGCOMM '20}
}

@misc{chakraborty2020distributed,
      title={Distributed Routing in a Quantum Internet}, 
      author={Kaushik Chakraborty and Filip Rozpedek and Axel Dahlberg and Stephanie Wehner},
      year={2019},
      eprint={1907.11630},
      archivePrefix={arXiv},
      primaryClass={quant-ph},
      url={https://arxiv.org/abs/1907.11630}, 
}

@INPROCEEDINGS{sundaram2024distributed,
  author={Sundaram, Ranjani G. and Gupta, Himanshu and Ramakrishnan, C. R.},
  booktitle={2024 IEEE International Conference on Quantum Computing and Engineering (QCE)}, 
  title={Distributing Quantum Circuits with Minimum Circuit Execution Time over Quantum Networks}, 
  year={2024},
  volume={01},
  pages={1938-1948},
  publisher={IEEE},
  url={https://doi.org/10.1109/QCE60285.2024.00224},
  doi={10.1109/QCE60285.2024.00224}}

@article{monroe2014large,
  title = {Large-scale modular quantum-computer architecture with atomic memory and photonic interconnects},
  author = {Monroe, C. and Raussendorf, R. and Ruthven, A. and Brown, K. R. and Maunz, P. and Duan, L.-M. and Kim, J.},
  journal = {Phys. Rev. A},
  volume = {89},
  number = {2},
  pages = {022317},
  numpages = {16},
  year = {2014},
  month = {Feb},
  publisher = {American Physical Society},
  doi = {10.1103/PhysRevA.89.022317},
  url = {https://link.aps.org/doi/10.1103/PhysRevA.89.022317}
}

@article{Gidney2021howtofactorbit,
  doi = {10.22331/q-2021-04-15-433},
  url = {https://doi.org/10.22331/q-2021-04-15-433},
  title = {How to factor 2048 bit {RSA} integers in 8 hours using 20 million noisy qubits},
  author = {Gidney, Craig and Eker{\aa{}}, Martin},
  journal = {{Quantum}},
  issn = {2521-327X},
  publisher = {{Verein zur F{\"{o}}rderung des Open Access Publizierens in den Quantenwissenschaften}},
  volume = {5},
  pages = {433},
  month = apr,
  year = {2021}
}

@article{cuomo2020towards,
author = {Cuomo, Daniele and Caleffi, Marcello and Cacciapuoti, Angela Sara},
title = {Towards a distributed quantum computing ecosystem},
year = {2020},
issue_date = {July 2020},
publisher = {John Wiley \& Sons, Inc.},
address = {USA},
volume = {1},
number = {1},
url = {https://doi.org/10.1049/iet-qtc.2020.0002},
doi = {10.1049/iet-qtc.2020.0002},
journal = {IET Quantum Communication},
month = jul,
pages = {3–8},
numpages = {6}
}

@inproceedings{rahman2026harvest,
  author    = {Mahir Rahman and Samuel Joseph and Nihar Kodkani and Behnaz Arzani and Vamsi Addanki},
  title     = {Harvest: Adaptive Photonic Switching Schedules for Collective Communication in Scale-up Domains},
  booktitle = {Proceedings of the ACM SIGCOMM 2026 Conference},
  series    = {SIGCOMM '26},
  year      = {2026},
  address   = {New York, NY, USA},
  location  = {Denver, CO, USA},
  pages     = {1576-1600},
  publisher = {Association for Computing Machinery},
  doi       = {10.1145/3789240.3829166}
}

@misc{pouryousef2026benchmarking,
  title = {Benchmarking Quantum Data Center Architectures: A Performance and Scalability Perspective},
  author = {Pouryousef, Shahrooz and Kaur, Eneet and Shapourian, Hassan and Towsley, Don and Kompella, Ramana and Nejabati, Reza},
  year = {2026},
  eprint = {2601.01353},
  archivePrefix = {arXiv},
  primaryClass = {quant-ph},
  url = {https://arxiv.org/abs/2601.01353}
}

@article{vanleent2022entangling,
  author  = {{van Leent}, Tim and Bock, Matthias and Fertig, Florian and Garthoff, Robert and
             Eppelt, Sebastian and Zhou, Yiru and Malik, Pooja and Seubert, Matthias and
             Bauer, Tobias and Rosenfeld, Wenjamin and Zhang, Wei and Becher, Christoph and
             Weinfurter, Harald},
  title   = {Entangling single atoms over 33\,km telecom fibre},
  journal = {Nature},
  volume  = {607},
  number  = {7917},
  pages   = {69--73},
  year    = {2022},
  doi     = {10.1038/s41586-022-04764-4},
  note    = {Polarization-preserving QFC of 780\,nm Rb photons to 1517\,nm with
             57\% external device efficiency per converter (2.44\,dB); 780\,nm fiber
             attenuates by a factor of 10 over 2.5\,km (4\,dB/km)}
}

@misc{zhao2026universal,
      title={A Universal Quantum Information Preserving Photonic Switch for Scalable Quantum Networks}, 
      author={Jiapeng Zhao and Stéphane Vinet and Amir Minoofar and Michael Kilzer and Lucas Wang and Galan Moody and Vijoy Pandey and Ramana Kompella and Reza Nejabati},
      year={2026},
      eprint={2604.21902},
      archivePrefix={arXiv},
      primaryClass={quant-ph},
      url={https://arxiv.org/abs/2604.21902}, 
}

@misc{keysightoptical,
  author = {{Keysight Technologies}},
  title  = {Automation Optical Switches},
  url    = {https://www.keysight.com/us/en/products/ethernet-traffic-emulation/lab-test-automation/automation-optical-switches.html},
  year = {2026}
}

@misc{amponsah2026tetris,
  author        = {Eliezer Amponsah and Deeksha P Rao and Vamsi Addanki},
  title         = {{Tetris}: Circuit Scheduling for Rearrangeably Non-Blocking Photonic Interconnects},
  year          = {2026},
  eprint        = {2609.25434},
  archivePrefix = {arXiv},
  primaryClass  = {cs.NI},
  url           = {https://arxiv.org/abs/2609.25434}
}

@inproceedings{photonicsreconfiguration,
  author = {Rao, Deeksha P and Addanki, Vamsi},
  title = {What Is Reconfiguration Delay, Really? An End-to-End View of Photonic Switching},
  year = {2026},
  booktitle = {Proceedings of the 25th ACM Workshop on Hot Topics in Networks (HotNets '26)}
}

@article{PhysRevA.100.032308,
  title = {Automated distribution of quantum circuits via hypergraph partitioning},
  author = {Andr{\'e}s-Mart{\'{\i}}nez, Pablo and Heunen, Chris},
  journal = {Phys. Rev. A},
  volume = {100},
  issue = {3},
  pages = {032308},
  numpages = {11},
  year = {2019},
  month = {Sep},
  publisher = {American Physical Society},
  doi = {10.1103/PhysRevA.100.032308},
  url = {https://link.aps.org/doi/10.1103/PhysRevA.100.032308}
}

@article{Andres-Martinez_2024,
doi = {10.1088/2058-9565/ad6734},
url = {https://doi.org/10.1088/2058-9565/ad6734},
year = {2024},
month = {aug},
publisher = {IOP Publishing},
volume = {9},
number = {4},
pages = {045021},
author = {Andres-Martinez, Pablo and Forrer, Tim and Mills, Daniel and Wu, Jun-Yi and Henaut, Luciana and Yamamoto, Kentaro and Murao, Mio and Duncan, Ross},
title = {Distributing circuits over heterogeneous, modular quantum computing network architectures},
journal = {Quantum Science and Technology}
}

@inproceedings{nejabati2022dynamic,
  author    = {Nejabati, Reza and Wang, Rui and Simeonidou, Dimitra},
  title     = {Dynamic Quantum Network: from Quantum Data Centre to Quantum Cloud Computing},
  booktitle = {Optical Fiber Communication Conference (OFC) 2022},
  publisher = {Optica Publishing Group},
  year      = {2022},
  pages     = {Th3D.1},
  doi       = {10.1364/OFC.2022.Th3D.1},
  url       = {https://opg.optica.org/abstract.cfm?uri=OFC-2022-Th3D.1}
}

@article{vista2025entanglement,
  author    = {Vista, Francesco and DiAdamo, Stephen and Kaur, Eneet and Shapourian, Hassan and Nejabati, Reza},
  title     = {Entanglement Request Scheduling in Quantum Datacenter Networks},
  journal   = {IEEE Network},
  volume    = {39},
  number    = {3},
  pages     = {188-195},
  year      = {2025},
  publisher = {IEEE},
  doi       = {10.1109/MNET.2025.3532847},
  url       = {https://doi.org/10.1109/MNET.2025.3532847}
}

@misc{pouryousef2025networkaware,
  author        = {Shahrooz Pouryousef and Reza Nejabati and Don Towsley and Ramana Kompella and Eneet Kaur},
  title         = {Network-Aware Scheduling for Remote Gate Execution in Quantum Data Centers},
  year          = {2025},
  eprint        = {2504.20176},
  archivePrefix = {arXiv},
  primaryClass  = {quant-ph},
  url           = {https://arxiv.org/abs/2504.20176}
}

@article{zhao2016benes,
  author  = {Zhao, Shuoyi and Lu, Liangjun and Zhou, Linjie and Li, Dong and Guo, Zhanzhi and Chen, Jianping},
  title   = {16 $\times$ 16 silicon {Mach--Zehnder} interferometer switch actuated with waveguide microheaters},
  journal = {Photonics Research},
  volume  = {4},
  number  = {5},
  pages   = {202--207},
  year    = {2016},
  doi     = {10.1364/PRJ.4.000202}
}

@article{zhao2025scalable,
  author  = {Zhao, Jiapeng and Xu, Yang and Lu, Xiyuan and Kaur, Eneet and Kilzer, Michael and Kompella, Ramana and Boyd, Robert W. and Nejabati, Reza},
  title   = {Scalable low-latency entanglement distribution for distributed quantum computing},
  journal = {Optica Quantum},
  volume  = {3},
  number  = {6},
  pages   = {606--616},
  year    = {2025},
  doi     = {10.1364/OPTICAQ.569352}
}

@INPROCEEDINGS{davis2023towards,
  author={Davis, Marc G. and Chung, Joaquin and Englund, Dirk and Kettimuthu, Rajkumar},
  booktitle={2023 IEEE International Conference on Quantum Computing and Engineering (QCE)}, 
  title={Towards Distributed Quantum Computing by Qubit and Gate Graph Partitioning Techniques}, 
  year={2023},
  volume={01},
  number={},
  pages={161-167},
  doi={10.1109/QCE57702.2023.00026}}

@article{wu2021sequence,
   title={SeQUeNCe: a customizable discrete-event simulator of quantum networks},
   volume={6},
   ISSN={2058-9565},
   url={http://dx.doi.org/10.1088/2058-9565/ac22f6},
   DOI={10.1088/2058-9565/ac22f6},
   number={4},
   journal={Quantum Science and Technology},
   publisher={IOP Publishing},
   author={Wu, Xiaoliang and Kolar, Alexander and Chung, Joaquin and Jin, Dong and Zhong, Tian and Kettimuthu, Rajkumar and Suchara, Martin},
   year={2021},
   month=Sept, pages={045027} }

@article{Main2025,
  author  = {Main, D. and Drmota, P. and Nadlinger, D. P. and
             Ainley, E. M. and Agrawal, A. and Nichol, B. C. and
             Srinivas, R. and Araneda, G. and Lucas, D. M.},
  title   = {Distributed quantum computing across an optical network link},
  journal = {Nature},
  year    = {2025},
  volume  = {638},
  number  = {8050},
  pages   = {383--388},
  doi     = {10.1038/s41586-024-08404-x},
  url     = {https://doi.org/10.1038/s41586-024-08404-x},
  issn    = {1476-4687}
}

@ARTICLE{10521772,
  author={Liu, Junyu and Jiang, Liang},
  journal={IEEE Network}, 
  title={Quantum Data Center: Perspectives}, 
  year={2024},
  volume={38},
  number={5},
  pages={160-166},
  doi={10.1109/MNET.2024.3397836}}

@article{Cacciapuoti_2026,
   title={Quantum Data Centres: why entanglement changes everything},
   volume={384},
   ISSN={1471-2962},
   url={http://dx.doi.org/10.1098/rsta.2024.0518},
   DOI={10.1098/rsta.2024.0518},
   number={2315},
   journal={Philosophical Transactions of the Royal Society A: Mathematical, Physical and Engineering Sciences},
   publisher={The Royal Society},
   author={Cacciapuoti, Angela Sara and Pellitteri, Claudio and Illiano, Jessica and d’Avossa, Laura and Mazza, Francesco and Chen, Siyi and Caleffi, Marcello},
   year={2026},
   month=Feb }

@article{BIRLICA2026108597,
title = {Entanglement in the trees: Optimal entanglement distribution in binary tree quantum network topologies},
journal = {Future Generation Computer Systems},
volume = {184},
pages = {108597},
year = {2026},
issn = {0167-739X},
doi = {https://doi.org/10.1016/j.future.2026.108597},
url = {https://www.sciencedirect.com/science/article/pii/S0167739X26002311},
author = {Iulian Ioan Bîrlică and Alin Bogdan Popa and Bogdan Călin Ciobanu and Florin Pop and Pantelimon George Popescu}
}

@article{quetschlich2023mqtbench,
  title        = {{{MQT Bench}}: {Benchmarking Software and Design Automation Tools for Quantum Computing}},
  shorttitle   = {{MQT Bench}},
  author       = {Quetschlich, Nils and Burgholzer, Lukas and Wille, Robert},
  year         = {2023},
  journal      = {{Quantum}},
  volume       = {7},
  pages        = {1062},
  doi          = {10.22331/q-2023-07-20-1062},
  note         = {{{MQT Bench}} is available at \url{https://mqt-bench.app/}},
  eprint       = {2204.13719},
  eprinttype   = {arxiv}
}

@article{CALEFFI2024110672,
title = {Distributed quantum computing: A survey},
journal = {Computer Networks},
volume = {254},
pages = {110672},
year = {2024},
issn = {1389-1286},
doi = {https://doi.org/10.1016/j.comnet.2024.110672},
url = {https://www.sciencedirect.com/science/article/pii/S1389128624005048},
author = {Marcello Caleffi and Michele Amoretti and Davide Ferrari and Jessica Illiano and Antonio Manzalini and Angela Sara Cacciapuoti}
}

@article{PRXQuantum,
  title = {Development of Quantum Interconnects (QuICs) for Next-Generation Information Technologies},
  author = {Awschalom, David and Berggren, Karl K. and Bernien, Hannes and Bhave, Sunil and Carr, Lincoln D. and Davids, Paul and Economou, Sophia E. and Englund, Dirk and Faraon, Andrei and Fejer, Martin and Guha, Saikat and Gustafsson, Martin V. and Hu, Evelyn and Jiang, Liang and Kim, Jungsang and Korzh, Boris and Kumar, Prem and Kwiat, Paul G. and Lon{\ifmmode \check{c}\else \v{c}\fi{}}ar, Marko and Lukin, Mikhail D. and Miller, David A.B. and Monroe, Christopher and Nam, Sae Woo and Narang, Prineha and Orcutt, Jason S. and Raymer, Michael G. and Safavi-Naeini, Amir H. and Spiropulu, Maria and Srinivasan, Kartik and Sun, Shuo and Vu{\ifmmode \check{c}\else \v{c}\fi{}}kovi{\ifmmode \acute{c}\else \'{c}\fi{}}, Jelena and Waks, Edo and Walsworth, Ronald and Weiner, Andrew M. and Zhang, Zheshen},
  journal = {PRX Quantum},
  volume = {2},
  issue = {1},
  pages = {017002},
  numpages = {21},
  year = {2021},
  month = {Feb},
  publisher = {American Physical Society},
  doi = {10.1103/PRXQuantum.2.017002},
  url = {https://link.aps.org/doi/10.1103/PRXQuantum.2.017002}
}

@article{c7bc8b77-97a9-3229-bba6-29cbf867d990,
 ISSN = {13645021, 14712946},
 URL = {http://www.jstor.org/stable/53167},
 author = {Christof Zalka},
 journal = {Proceedings: Mathematical, Physical and Engineering Sciences},
 number = {1969},
 pages = {313--322},
 publisher = {Royal Society},
 title = {Simulating Quantum Systems on a Quantum Computer},
 urldate = {2026-11-17},
 volume = {454},
 year = {1998}
}

@article{doi:10.1137/S0097539795293172,
author = {Shor, Peter W.},
title = {Polynomial-Time Algorithms for Prime Factorization and Discrete Logarithms on a Quantum Computer},
journal = {SIAM Journal on Computing},
volume = {26},
number = {5},
pages = {1484-1509},
year = {1997},
doi = {10.1137/S0097539795293172},
URL = {https://doi.org/10.1137/S0097539795293172},
eprint = {https://doi.org/10.1137/S0097539795293172}
}

@article{10.1145/3674151,
author = {Ang, James and Carini, Gabriella and Chen, Yanzhu and Chuang, Isaac and Demarco, Michael and Economou, Sophia and Eickbusch, Alec and Faraon, Andrei and Fu, Kai-Mei and Girvin, Steven and Hatridge, Michael and Houck, Andrew and Hilaire, Paul and Krsulich, Kevin and Li, Ang and Liu, Chenxu and Liu, Yuan and Martonosi, Margaret and McKay, David and Misewich, Jim and Ritter, Mark and Schoelkopf, Robert and Stein, Samuel and Sussman, Sara and Tang, Hong and Tang, Wei and Tomesh, Teague and Tubman, Norm and Wang, Chen and Wiebe, Nathan and Yao, Yongxin and Yost, Dillon and Zhou, Yiyu},
title = {ARQUIN: Architectures for Multinode Superconducting Quantum Computers},
year = {2024},
issue_date = {September 2024},
publisher = {Association for Computing Machinery},
address = {New York, NY, USA},
volume = {5},
number = {3},
url = {https://doi.org/10.1145/3674151},
doi = {10.1145/3674151},
journal = {ACM Transactions on Quantum Computing},
month = sep,
articleno = {19},
numpages = {59}
}

@misc{tekleab2026distributedswitchingprotocolquantum,
      title={A Distributed Switching Protocol for Quantum Networks}, 
      author={Aman Yacob Tekleab and Yifeng Shen and Yoshii Yutaro and Amin Taherkhani and Rodney Van Meter and Shota Nagayama},
      year={2026},
      eprint={2605.07295},
      archivePrefix={arXiv},
      primaryClass={quant-ph},
      url={https://arxiv.org/abs/2605.07295}, 
}

@inproceedings{Koyama_2024,
   title={Optimal Switching Networks for Paired-Egress Bell State Analyzer Pools},
   url={http://dx.doi.org/10.1109/QCE60285.2024.00219},
   DOI={10.1109/qce60285.2024.00219},
   booktitle={2024 IEEE International Conference on Quantum Computing and Engineering (QCE)},
   publisher={IEEE},
   author={Koyama, Marii and Yun, Claire and Taherkhani, Amin and Benchasattabuse, Naphan and Sane, Bernard Ousmane and Hajdušek, Michal and Nagayama, Shota and Meter, Rodney Van},
   year={2024},
   month=Sept, pages={1897–1907} }

@ARTICLE{9334411,
  author={Ferrari, Davide and Cacciapuoti, Angela Sara and Amoretti, Michele and Caleffi, Marcello},
  journal={IEEE Transactions on Quantum Engineering}, 
  title={Compiler Design for Distributed Quantum Computing}, 
  year={2021},
  volume={2},
  number={},
  pages={1-20},
  doi={10.1109/TQE.2021.3053921}}

@misc{earl2022architecturefirstgenerationcommercialquantum,
      title={Architecture of a First-Generation Commercial Quantum Network}, 
      author={Duncan Earl and K Karunaratne and Jason Schaake and Ryan Strum and Patrick Swingle and Ryan Wilson},
      year={2022},
      eprint={2211.14871},
      archivePrefix={arXiv},
      primaryClass={quant-ph},
      url={https://arxiv.org/abs/2211.14871}, 
}

@INPROCEEDINGS{10628181,
  author={Mazza, Francesco and Caleffi, Marcello and Cacciapuoti, Angela Sara},
  booktitle={2024 International Conference on Quantum Communications, Networking, and Computing (QCNC)}, 
  title={Quantum LAN: On-Demand Network Topology via Two-colorable Graph States}, 
  year={2024},
  volume={},
  number={},
  pages={127-134},
  doi={10.1109/QCNC62729.2024.00029}}

@inproceedings{Mazza_2025,
   title={Simulation of Entanglement-Enabled Connectivity in Qlans Using SeQUeNCe},
   url={http://dx.doi.org/10.1109/ICC52391.2025.11434940},
   DOI={10.1109/icc52391.2025.11434940},
   booktitle={ICC 2025 - IEEE International Conference on Communications},
   publisher={IEEE},
   author={Mazza, Francesco and Zhan, Caitao and Chung, Joaquin and Kettimuthu, Rajkumar and Caleffi, Marcello and Cacciapuoti, Angela Sara},
   year={2025},
   month=June, pages={1–6} }

@misc{blanco2025physicallogicalgraphstatebasedconnectivity,
      title={From Physical to Logical: Graph-State-Based Connectivity in Quantum Networks}, 
      author={Mateo M. Blanco and Manuel Fernández-Veiga and Ana Fernández-Vilas and Rebeca P. Díaz-Redondo},
      year={2025},
      eprint={2509.08384},
      archivePrefix={arXiv},
      primaryClass={quant-ph},
      url={https://arxiv.org/abs/2509.08384}, 
}

@INPROCEEDINGS{9796814,
  author={Zhao, Yangming and Zhao, Gongming and Qiao, Chunming},
  booktitle={IEEE INFOCOM 2022 - IEEE Conference on Computer Communications}, 
  title={E2E Fidelity Aware Routing and Purification for Throughput Maximization in Quantum Networks}, 
  year={2022},
  volume={},
  number={},
  pages={480-489},
  doi={10.1109/INFOCOM48880.2022.9796814}}

@Article{Abbas2024,
author={Abbas, Amira
and Ambainis, Andris
and Augustino, Brandon
and B{\"a}rtschi, Andreas
and Buhrman, Harry
and Coffrin, Carleton
and Cortiana, Giorgio
and Dunjko, Vedran
and Egger, Daniel J.
and Elmegreen, Bruce G.
and Franco, Nicola
and Fratini, Filippo
and Fuller, Bryce
and Gacon, Julien
and Gonciulea, Constantin
and Gribling, Sander
and Gupta, Swati
and Hadfield, Stuart
and Heese, Raoul
and Kircher, Gerhard
and Kleinert, Thomas
and Koch, Thorsten
and Korpas, Georgios
and Lenk, Steve
and Marecek, Jakub
and Markov, Vanio
and Mazzola, Guglielmo
and Mensa, Stefano
and Mohseni, Naeimeh
and Nannicini, Giacomo
and O'Meara, Corey
and Tapia, Elena Pe{\~{n}}a
and Pokutta, Sebastian
and Proissl, Manuel
and Rebentrost, Patrick
and Sahin, Emre
and Symons, Benjamin C. B.
and Tornow, Sabine
and Valls, V{\'i}ctor
and Woerner, Stefan
and Wolf-Bauwens, Mira L.
and Yard, Jon
and Yarkoni, Sheir
and Zechiel, Dirk
and Zhuk, Sergiy
and Zoufal, Christa},
title={Challenges and opportunities in quantum optimization},
journal={Nature Reviews Physics},
year={2024},
month={Dec},
day={01},
volume={6},
number={12},
pages={718-735},
issn={2522-5820},
doi={10.1038/s42254-024-00770-9},
url={https://doi.org/10.1038/s42254-024-00770-9}
}

@article{Li_2016,
   title={Efficient and low-noise single-photon-level frequency conversion interfaces using silicon nanophotonics},
   volume={10},
   ISSN={1749-4893},
   url={http://dx.doi.org/10.1038/nphoton.2016.64},
   DOI={10.1038/nphoton.2016.64},
   number={6},
   journal={Nature Photonics},
   publisher={Springer Science and Business Media LLC},
   author={Li, Qing and Davanço, Marcelo and Srinivasan, Kartik},
   year={2016},
   month=Apr, pages={406–414} }

@article{RevModPhys.86.153,
  title = {Quantum simulation},
  author = {Georgescu, I. M. and Ashhab, S. and Nori, Franco},
  journal = {Rev. Mod. Phys.},
  volume = {86},
  issue = {1},
  pages = {153--185},
  numpages = {33},
  year = {2014},
  month = {Mar},
  publisher = {American Physical Society},
  doi = {10.1103/RevModPhys.86.153},
  url = {https://link.aps.org/doi/10.1103/RevModPhys.86.153}
}

@misc{farhi2014qaoa,
      title={A Quantum Approximate Optimization Algorithm}, 
      author={Edward Farhi and Jeffrey Goldstone and Sam Gutmann},
      year={2014},
      eprint={1411.4028},
      archivePrefix={arXiv},
      primaryClass={quant-ph},
      url={https://arxiv.org/abs/1411.4028}, 
}

@Article{Daley2022,
author={Daley, Andrew J.
and Bloch, Immanuel
and Kokail, Christian
and Flannigan, Stuart
and Pearson, Natalie
and Troyer, Matthias
and Zoller, Peter},
title={Practical quantum advantage in quantum simulation},
journal={Nature},
year={2022},
month={Jul},
day={01},
volume={607},
number={7920},
pages={667-676},
issn={1476-4687},
doi={10.1038/s41586-022-04940-6},
url={https://doi.org/10.1038/s41586-022-04940-6}
}

@article{Montanaro2016,
  title={Quantum algorithms: an overview},
  author={Ashley Montanaro},
  journal={npj Quantum Information},
  year={2015},
  volume={2},
  url={https://api.semanticscholar.org/CorpusID:2992738}
}

@INPROCEEDINGS{365700,
  author={Shor, P.W.},
  booktitle={Proceedings 35th Annual Symposium on Foundations of Computer Science}, 
  title={Algorithms for quantum computation: discrete logarithms and factoring}, 
  year={1994},
  volume={},
  number={},
  pages={124-134},
  doi={10.1109/SFCS.1994.365700}}

@misc{promponas2026ratefidelitytradeoffsallphotonicmemoryequipped,
      title={Rate-Fidelity Tradeoffs in All-Photonic and Memory-Equipped Quantum Switches}, 
      author={Panagiotis Promponas and Leonardo Bacciottini and Paul Polakos and Gayane Vardoyan and Don Towsley and Leandros Tassiulas},
      year={2026},
      eprint={2603.02610},
      archivePrefix={arXiv},
      primaryClass={quant-ph},
      url={https://arxiv.org/abs/2603.02610}, 
}

@misc{piparo2026adaptiveresourcememorycontrol,
      title={Adaptive Resource and Memory Control for Stability in Quantum Entanglement Distribution}, 
      author={Nicolò Lo Piparo and William J. Munro and Kae Nemoto},
      year={2026},
      eprint={2603.24874},
      archivePrefix={arXiv},
      primaryClass={quant-ph},
      url={https://arxiv.org/abs/2603.24874}, 
}

@misc{elsayed2026balancingquantummemoriesasymmetric,
      title={Balancing Quantum Memories in Asymmetric Repeaters for High-Fidelity Entanglement Distribution}, 
      author={Karim S. Elsayed and Amr Rizk},
      year={2026},
      eprint={2604.24554},
      archivePrefix={arXiv},
      primaryClass={quant-ph},
      url={https://arxiv.org/abs/2604.24554}, 
}

@article{Rowland_2025,
   title={Characterization of near-infrared to telecom frequency conversion in a rubidium-filled hollow-core photonic-crystal fiber},
   volume={33},
   ISSN={1094-4087},
   url={http://dx.doi.org/10.1364/OE.549459},
   DOI={10.1364/oe.549459},
   number={8},
   journal={Optics Express},
   publisher={Optica Publishing Group},
   author={Rowland, Jed A. and Perrella, Christopher and Offer, Rachel F. and Luiten, Andre N. and Sparkes, Ben M. and Weinhold, Till J.},
   year={2025},
   month=Apr, pages={18076} }

\end{document}